\documentclass[a4paper,fleqn,usenatbib]{mnras}

\usepackage{newtxtext,newtxmath}

\usepackage[T1]{fontenc}
\usepackage{ae,aecompl}
\usepackage{bm}
\usepackage{xcolor}

\usepackage{float}
\usepackage[export]{adjustbox}
\usepackage{graphicx}	

\usepackage{amsmath}	
\usepackage{amssymb}	

\usepackage{multicol}
\usepackage{verbatim}   
\usepackage{threeparttable}
\usepackage{upgreek}
\usepackage{rotating}

\usepackage{subfig}
\usepackage{comment}
\usepackage{longtable}
\usepackage{caption}
\usepackage{threeparttable}
\newcommand{\lcdm}{$\Lambda$CDM}

\newcommand{\om}{\Omega_{m0}}
\newcommand{\ol}{\Omega_{\Lambda}}
\newcommand{\ok}{\Omega_{k0}}
\newcommand{\FT}[1]{}

\definecolor{MZ}{RGB}{255,128,0}

\title[Constraints from Mg II QSO data]{Standardizing reverberation-measured Mg II time-lag quasars, by using the radius-luminosity relation, and constraining cosmological model parameters}

\author[]{
Narayan Khadka$^{1}$\thanks{E-mail: nkhadka@phys.ksu.edu},
Zhefu Yu$^{2}$,
Michal Zaja\v{c}ek$^{3,4}$,
Mary Loli Martinez-Aldama$^3$,
\newauthor \hspace{0.1mm}
Bo\.{z}ena Czerny$^3$,
Bharat Ratra$^{1}$\thanks{E-mail: ratra@phys.ksu.edu}\\
$^{1}$Department of Physics, Kansas State University, 116 Cardwell Hall, Manhattan, KS 66502, USA\\
$^{2}$Department of Astronomy, The Ohio State University, Columbus, OH 43210, USA\\
$^{3}$Center for Theoretical Physics, Polish Academy of Sciences, Al.\ Lotnik\'{o}w 32/46, 02-668 Warsaw, Poland\\
$^{4}$Department of Theoretical Physics and Astrophysics, Faculty of Science, Masaryk University, Kotl\'a\v{r}sk\'a 2, 611 37 Brno, Czech Republic
}

\date{Accepted XXX. Received YYY; in original form ZZZ}

\pubyear{2019}

\begin{document}
\label{firstpage}
\pagerange{\pageref{firstpage}--\pageref{lastpage}}
\maketitle

\begin{abstract}
We use 78 reverberation-measured Mg II time-lag quasars (QSOs) in the redshift range $0.0033 \leq z \leq 1.89$ to constrain cosmological parameters in six different cosmological models. The basis of our method is the use of the radius-luminosity or $R-L$ relation to standardize these 78 Mg II QSOs. In each cosmological model we simultaneously determine $R-L$ relation and cosmological model parameters, thus avoiding the circularity problem. We find that the $R-L$ relation parameter values are independent of the cosmological model used in the analysis thus establishing that current Mg II QSOs are standardizable candles. Cosmological constraints obtained using these QSOs are significantly weaker than, but consistent with, those obtained from a joint analysis of baryon acoustic oscillation (BAO) observations and Hubble parameter [$H(z)$] measurements. So, we also analyse these QSOs in conjunction with the BAO + $H(z)$ data and find cosmological constraints consistent with the standard spatially-flat $\Lambda$CDM model as well as with mild dark energy dynamics and a little spatial curvature. A larger sample of higher-quality reverberation-measured QSOs should have a smaller intrinsic dispersion and so should provide tighter constraints on cosmological parameters.
\end{abstract}




\section{Introduction}
\label{sec:Introduction}

It is a well-established fact that our Universe is currently undergoing accelerated cosmological expansion \citep{Farooqetal2017, Scolnicetal2018, PlanckCollaboration2020, eBOSSCollaboration2021}. This observational fact can be explained by general relativistic cosmological models if we include dark energy in them. The simplest cosmological model that is consistent with this observation is the standard spatially-flat $\Lambda$CDM model \citep{Peebles1984}. In this model, dark energy in the form of the cosmological constant $\Lambda$ contributes $\sim 70\%$ of the current cosmological energy budget, non-relativistic cold dark matter (CDM) contributes $\sim 25\%$, and almost all of the remaining $\sim 5\%$ is contributed by non-relativistic baryons. This model is consistent with most observational data but a little spatial curvature and mild dark energy dynamics are not ruled out. So, in this paper, in addition to the $\Lambda$CDM model, we consider two dynamical dark energy models, one being the widely-used but physically-incomplete XCDM parametrization which parametrizes dynamical dark energy as an $X$-fluid and the other is the physically-complete $\phi$CDM model which models dynamical dark energy as a scalar field. In each case we consider flat and non-flat spatial hypersurfaces to also allow for possibly non-zero spatial curvature of the Universe.\footnote{Recent observational constraints on spatial curvature are discussed in \citet{Farooqetal2015}, \citet{Chenetal2016}, \citet{Ranaetal2017}, \citet{Oobaetal2018a, Oobaetal2018b}, \citet{Yuetal2018}, \citet{ParkRatra2019a, ParkRatra2019b}, \citet{Wei2018}, \citet{DESCollaboration2019}, \citet{Lietal2020}, \citet{Handley2019}, \citet{EfstathiouGratton2020}, \citet{DiValentinoetal2021a}, \citet{VelasquezToribioFabris2020}, \citet{Vagnozzietal2020, Vagnozzietal2021}, \citet{KiDSCollaboration2021}, \citet{ArjonaNesseris2021}, \citet{Dhawanetal2021}, and references therein.}

These models are mostly tested using well-established cosmological probes such as cosmic microwave background (CMB) anisotropy data, baryon acoustic oscillation (BAO) observations, Hubble parameter [$H(z)$] measurements, and Type Ia supernova (SNIa) apparent magnitude data. CMB anisotropy data probe the $z \sim 1100$ part of redshift space and are the only high-redshift data. BAO data probe redshift space up to $z \sim 2.3$, the highest $z$ reached by the better-established lower-redshift probes. These are limited sets of cosmological data and a number of observationally-viable cosmological models make very similar predictions for these probes, so to establish a more accurate standard cosmological model and to obtain tighter cosmological parameter constraints we need to use other astronomical data.

A significant amount of work has been done to develop new cosmological probes. This work includes use of HII starburst galaxy observations which extend to $z \sim 2.4$ \citep{ManiaRatra2012, Chavezetal2014, GonzalezMoran2019, GonzalezMoranetal2021, Caoetal2020, Caoetal2021a, Johnsonetal2021}, quasar (QSO) angular size measurements which extend to $z \sim 2.7$ \citep{Caoetal2017, Ryanetal2019, Caoetal2020, Caoetal2021b, Zhengetal2021, Lianetal2021}, QSO X-ray and UV flux measurements which extend to $z \sim 7.5$ \citep{RisalitiLusso2015, RisalitiLusso2019, KhadkaRatra2020a, KhadkaRatra2020b, KhadkaRatra2021a, KhadkaRatra2021b, Yangetal2020, Lussoetal2020, Lietal2021, Lianetal2021}, and gamma-ray burst (GRB) data that extend to $z \sim 8.2$ \citep{Amati2008, Amati2019, samushia_ratra_2010, Wang_2016, Demianski_2019, Dirirsa2019, KhadkaRatra2020c, Khadkaetal2021}.

An additional new method that can be used in cosmology is based on QSOs with a measured time delay between the quasar ionizing continuum and the Mg II line luminosity. This technique is referred to as reverberation mapping and it makes use of the tight correlation between the variable ionizing radiation powered by the accretion disc and the line-emission that originates in the broad-line region (BLR) optically-thick material located farther away that efficiently reprocesses the disc continuum radiation \citep{1982ApJ...255..419B}. We refer to these reverberation-mapped sources as Mg II QSOs. We use Mg II QSOs to constrain cosmological dark energy models for the following reasons: (i) The current reasonably large number, 78, of studied Mg II QSOs at intermediate $z$ \citep{2019ApJ...880...46C,Homayouni2020, Mary2020, Michal2020, Michal2021, Zhefu2021}. The current Mg II QSO redshift range $0.0033\leq z \leq 1.89$ is more extended, especially towards higher redshifts, than that of $117$ reverberation-mapped H$\beta$ quasars \citep[$0.002\leq z \leq 0.89$; ][]{Mary2019}. (ii) Some works using QSO X-ray and UV flux measurements show evidence for tension with predictions of the standard spatially-flat $\Lambda$CDM model with $\Omega_{m0}=0.3$ \citep{RisalitiLusso2019, KhadkaRatra2020b, KhadkaRatra2021a, KhadkaRatra2021b, Lussoetal2020} and the Mg II QSO sample is an alternative QSO data set that might help clarify this issue. (iii) For MgII QSOs, the UV spectrum is not severely contaminated by starlight as is the case of QSOs where reverberation mapping has been performed using the optical H$\beta$ line \citep{Bentz2013}. Hence, the measured Mg II QSO flux density at 3000\,\AA\, can be considered to largely represent the accretion-disc ionizing flux density at this wavelength that is reprocessed by BLR clouds located at the mean distance of $R=c\tau$, where $\tau$ is the rest-frame time delay between the UV ionizing continuum and the broad-line material emitting Mg II inferred e.g. by the cross-correlation function.

The reveberation-measured rest-frame time-delay of the broad UV Mg II emission line (which is centered at 2798\,\AA\, in the rest frame) and the monochromatic luminosity of the QSO are correlated through the radius-luminosity correlation, also known as the $R-L$ relation, with the power-law scaling $R\propto L^{\gamma}$. Such a relation was first discovered for the broad H$\beta$ line in the optical domain \citep[the H$\beta$ rest-frame wavelength is 4860\,\AA;][]{kaspi2000,peterson2004,Bentz2013}, and the possibility of using such measurements to create a Hubble diagram and constrain cosmological parameters was discussed soon afterwards \citep{watson2011,haas2011,czerny2013,Bentz2013}. Using the H$\beta$ broad component, initially the power-law index $\gamma=0.67 \pm 0.05$ deviated from $\gamma=0.5$ given by simple photoionization arguments\footnote{Using the definition of the ionization parameter for a BLR cloud, $U=Q(H)/[4\pi  R^2 c n(H)]$, where $Q(H)$ is the hydrogen-ionizing photon flux in ${\rm cm^{-2} s^{-1}}$, $R$ is the cloud distance from the continuum source, and $n(H)$ is the total hydrogen density. Assuming that $Un(H)=\mathrm{const}$ for BLR clouds in different sources, we obtain $R\propto L^{1/2}$.} \citep{2005ApJ...629...61K}. After extending the sample by including lower-redshift sources and correcting for host starlight contamination \citep{Bentz2013}, the updated H$\beta$ sample yielded a slope of $\gamma=0.533^{+0.035}_{-0.033}$, i.e. consistent with the simple photoionization theory, and a small intrinsic scatter of only $\sigma_{\rm ext}=0.13$ dex, which made these data attractive for cosmological applications. As the $H\beta$ quasar sample was enlarged by adding sources with a higher accretion rate, the overall scatter increased significantly \citep{2014ApJ...782...45D,2018ApJ...856....6D,2017ApJ...851...21G}. Using accretion-rate tracers, such as the Eddington ratio, dimensionless accretion-rate, relative Fe II strength, or the fractional variability, it was found that this scatter is mostly driven by the accretion rate \citep{2018ApJ...856....6D,Mary2019,2020ApJ...903..112D}. Sources with a higher accretion rate have shortened time lags with respect to the $R-L$ relation, i.e. the higher the acrretion rate, the larger the departure. The same trend was later confirmed for the Mg II QSO $R-L$ relation \citep{Michal2020,Mary2020,Michal2021}. The deviation could also depend on the UV/optical SED or the amount of ionizing photons \citep{2020ApJ...899...73F}, which, however, is also linked directly or indirectly to the accretion rate via the thin accretion disc thermal SED, specifically the Big Blue Bump in the standard accretion theory \citep[BBB;][]{1987ApJ...321..305C,2019arXiv190106507K}.

The $R-L$ correlation, although with a relatively large dispersion of $\sim 0.3$ dex for Mg II QSOs \citep{Mary2020,Michal2021}, in principle enables us to use reverberation-measured Mg II QSOs to determine constraints on cosmological parameters since the time delay measurement allows one to obtain the source absolute luminosity \citep[see][ for overviews]{2019FrASS...6...75P,2020mbhe.confE..10M}. Some attempts have previously been made to use reverberation-measured QSOs in cosmology \citep{Mary2019,Czerny2021, Michal2021}, and so far an overall agreement has been found with the standard $\Lambda$CDM cosmological model for H$\beta$ QSOs \citep{Mary2019}, combined H$\beta$ and Mg II sources \citep{Czerny2021}, and Mg II QSOs alone \citep{Michal2021}.

In this paper, we use 78 Mg II QSOs --- the largest set of such measurements to date --- to simultaneously constrain cosmological parameters and $R-L$ relation parameters (the intercept $\beta$ and the slope $\gamma$) in six different cosmological models. This simultaneous determination of cosmological parameters and $R-L$ relation parameters --- done here for Mg II QSOs for the first time --- allows us to avoid the circularity problem. This is the problem of having to either assume $\beta$ and $\gamma$ to use the $R-L$ relation and data to constrain cosmological model parameters, or having to assume a cosmological model (and parameter values) to use the measurements to determine $\beta$ and $\gamma$. Since we determine $\beta$ and $\gamma$ values in six different cosmological models, we are able to test whether Mg II QSOs are standardizable candles.\footnote{This is one reason why we study a number of different cosmological models.} We find that the $R-L$ relation parameters are independent of the cosmological model in which they were derived, thus establishing that current Mg II QSOs are standardizable candles. However, while cosmological parameter constraints obtained using these Mg II QSOs are consistent with those obtained from most other cosmological probes, they are significantly less restrictive. The Mg II QSO constraints are less restrictive because the $R-L$ relation, which is the basis of our method, has a large intrinsic dispersion ($\sigma_{\rm ext} \sim 0.29$ dex) and also involves two nuisance parameters, $\beta$ and $\gamma$. Cosmological constraints obtained using the Mg II QSO data set are consistent with those obtained using BAO + $H(z)$ data, so we also analyze these 78 Mg II QSO data in conjunction with BAO + $H(z)$ data. Results obtained from the joint analyses are consistent with the standard spatially-flat $\Lambda$CDM model but also do not rule out a little spatial curvature and mild dark energy dynamics.

This paper is structured as follows. In Sec.\ 2 we summarize the cosmological models we use. In Sec. 3 we describe the data sets we analyze. In Sec. 4 we summarize our analysis methods. In Sec. 5 we present our results. We conclude in Sec. 6. The MgII QSO data sets we use are tabulated in the Appendix.

\section{Models}
\label{sec:models}

We constrain cosmological model parameters by comparing model predictions to cosmological measurements at known redshift $z$. We consider six different dark energy cosmological models, three with flat spatial geometry and three with non-flat spatial geometry. For the observations we consider, model predictions depend on the Hubble parameter --- the cosmological expansion rate --- a function that depends on $z$ and on the cosmological parameters of the model.

In these models the Hubble parameter can be expressed as
\begin{equation}
\label{eq:friedLCDM}
    H(z) = H_0\sqrt{\Omega_{m0}(1 + z)^3 + \Omega_{k0}(1 + z)^2 + \Omega_{DE}(z)},
\end{equation}
where $H_0$ is the Hubble constant, $\Omega_{DE}(z)$ is the dark energy density parameter, and $\Omega_{m0}$ and $\Omega_{k0}$ are the current values of the non-relativistic matter and curvature energy density parameters. In the spatially-flat models $\Omega_{k0} = 0$. For analyses of the BAO + $H(z)$ and QSO + BAO + $H(z)$ data, we express $\Omega_{m0}$ in terms of the current values of the cold dark matter density parameter $(\Omega_{c})$ and the baryon density parameter $(\Omega_{b})$: $\Omega_{m0} = \Omega_{c} + \Omega_{b}$, and use $\Omega_b h^2$ and $\Omega_c h^2$ as free parameters [here $h = H_0/(100\, {\rm km}\, {\rm s}^{-1} {\rm Mpc}^{-1}$)] instead of $\Omega_{m0}$. As discussed in Sec.\ 4, QSO data alone cannot constrain $H_0$, which in this case is set to $70$ ${\rm km}\hspace{1mm}{\rm s}^{-1}{\rm Mpc}^{-1}$; for the BAO + $H(z)$ and QSO + BAO + $H(z)$ data analyses $H_0$ is a free parameter to be determined from the data. The dark energy density evolves as a power of $(1+z)$ in four of the six models we study. In these models $\Omega_{DE}(z) = \Omega_{DE0}(1+z)^{1+\omega_X}$ where $\omega_X$ is the dark energy equation of state parameter (defined below) and $\Omega_{DE0}$ is the current value of the dark energy density parameter.

In the $\Lambda$CDM model $\omega_X = -1$ so $\Omega_{DE}$ = $\Omega_{DE0}$ = $\Omega_{\Lambda}$, and dark energy is the standard cosmological constant. The current values of the three $\Lambda$CDM model energy density parameters obey the energy budget equation $\Omega_{m0} + \Omega_{k0} + \Omega_{\Lambda} = 1$. For the QSO-only data analyses we fix $H_0$ and in the spatially-flat $\Lambda$CDM model we take $\Omega_{m0}$ to be the free parameter while in the non-flat $\Lambda$CDM model $\Omega_{m0}$, and $\Omega_{k0}$ are the free parameters.

In the XCDM parametrization dark energy is parametrized as an ideal $X$-fluid with equation of state parameter $\omega_X$ being the ratio of the $X$-fluid pressure and energy density. Here $\Omega_{DE0}$ = $\Omega_{X0}$ is the current value of the $X$-fluid dynamical dark energy density parameter. The current values of the three XCDM parametrization energy density parameters obey the energy budget equation $\Omega_{m0} + \Omega_{k0} + \Omega_{X0} = 1$. The $X$-fluid energy density decreases with time when $\omega_X > -1$. For the QSO-only data analyses we fix $H_0$ and in the spatially-flat XCDM parametrization we take $\Omega_{m0}$ and $\omega_X$ to be the free parameters while in the non-flat XCDM parametrization, $\Omega_{m0}$, $\Omega_{k0}$, and $\omega_X$ are the free parameters. In the limit $\omega_x \rightarrow -1$ the XCDM parametrization reduces to the $\Lambda$CDM model.

In the $\phi$CDM model \citep{PeeblesRatra1988, RatraPeebles1988, Pavlovetal2013} dynamical dark energy is a scalar field $\phi$.\footnote{Recent observational constraints on the $\phi$CDM model are discussed in \citet{Avsajanishvilietal2015}, \citet{SolaPeracaulaetal2018, SolaPercaulaetal2019}, \citet{Zhaietal2017}, \citet{Oobaetal2018c, Oobaetal2019}, \citet{ParkRatra2018, ParkRatra2019c, ParkRatra2020}, \citet{Sangwanetal2018}, \citet{Singhetal2019}, \citet{UrenaLopezRoy2020}, \citet{SinhaBanerjee2021}, and references therein.} Here the dynamical dark energy scalar field density parameter $\Omega_{DE}$ is determined by the potential energy density of the scalar field. In this paper we assume an inverse power law scalar field potential energy density
\begin{equation}
\label{eq:phiCDMV}
    V(\phi) = \frac{1}{2}\kappa m_{p}^2 \phi^{-\alpha}.
\end{equation}
In this equation $m_{p}$ is the Planck mass, $\alpha$ is a positive parameter [$\Omega_{DE}$ = $\Omega_{\phi}(z, \alpha)$ is the scalar field dynamical dark energy density parameter], and the constant $\kappa$ is determined by using the shooting method to ensure that the current energy budget constraint $\Omega_{m0} + \Omega_{k0} + \Omega_{\phi}(z = 0, \alpha) = 1$ is satisfied.

For this potential energy density, the equations of motion for a spatially homogeneous scalar field and FLRW metric tensor are
\begin{align}
\label{field}
   & \ddot{\phi} + 3\frac{\dot{a}}{a}\dot\phi - \frac{1}{2}\alpha \kappa m_{p}^2 \phi^{-\alpha - 1} = 0,  \\
\label{friedpCDM}
   & \left(\frac{\dot{a}}{a}\right)^2 = \frac{8 \uppi}{3 m_{p}^2}\left(\rho_m + \rho_{\phi}\right) - \frac{k}{a^2}.
\end{align}
Here $a$ is the scale factor, an overdot denotes a derivative with respect to time, $k$ is negative, zero, and positive for open, flat, and closed spatial geometries (corresponding to $\Omega_{k0} > 0, =0, \rm and < 0$), $\rho_m$ is the non-relativistic matter energy density, and the scalar field energy density
\begin{equation}
    \rho_{\phi} = \frac{m^2_p}{32\pi}[\dot{\phi}^2 + \kappa m^2_p \phi^{-\alpha}].
\end{equation}
We numerically integrate eqs.\ (3) and (4), compute $\rho_{\phi}$, and then compute $\Omega_{\phi}(z, \alpha)$ from
\begin{equation}
    \Omega_{\phi}(z, \alpha) = \frac{8 \uppi \rho_{\phi}}{3 m^2_p H^2_0}.
\end{equation}
For the QSO-only data analyses we fix $H_0$ and in the spatially-flat $\phi$CDM model we take $\Omega_{m0}$ and $\alpha$ to be the free parameters and in the non-flat $\phi$CDM model, $\Omega_{m0}$, $\Omega_{k0}$, and $\alpha$ are the free parameters. In the limit $\alpha\rightarrow0$ the $\phi$CDM model reduces to the $\Lambda$CDM model.

\section{Data}
\label{sec:data}

We use three different Mg II QSO compilations, as well as BAO and $H(z)$ data. The Mg II QSO data sets are summarized in Table \ref{tab:samples}, which lists the number of QSOs in each sample, and the covered redshift range. These data are listed in Table \ref{tab:MgQSOdata} where for each source the name, $z$, measured QSO flux for the Mg II line $(F_{3000})$, and rest-frame time-delay $(\tau)$ are listed.

\begin{table}
	\centering
	\caption{Summary of the Mg II QSO data sets.}
	\label{tab:samples}
	\begin{threeparttable}
	\begin{tabular}{l|cc}
	\hline
	Data set & Number & Redshift range \\
	\hline
	Mg II QSO-69 & $69$ & $[0.0033, 1.89]$\\
	Mg II QSO-9 & $9$ & $[1.06703, 1.7496]$\\
	Mg II QSO-78 & $78$ & $[0.0033, 1.89]$\\
	\hline
	\end{tabular}
    \end{threeparttable}
\end{table}

\begin{itemize}

\item[]{\bf Mg II QSO-69 sample}. This sample includes the first 69 QSOs listed in Table \ref{tab:MgQSOdata}. These data were originally analyzed and described in several publications. The Mg II QSO-69 sample contains 69 QSOs including those from the most recent Mg II Sloan Digital Sky Survey Reverberation Mapping data set \citep[SDSS-RM, 57 sources;][]{Homayouni2020}, from previous SDSS-RM results \citep[6 sources; ][where one source is included in the more recent SDSS-RM sample]{2016ApJ...818...30S}, several luminous quasars, in particular CTS 252 \citep{2018ApJ...865...56L}, CTS C30.10 \citep{2019ApJ...880...46C}, HE 0413-4031 \citep{Michal2020}, and HE 0435-4312 \citep{Michal2021}, and two older \textit{International Ultraviolet Explorer (IUE)} measurements of the low-luminosity QSO NGC 4151 based on two separate campaigns in 1988 and 1991 \citep{2006ApJ...647..901M}\footnote{Since there were two campaigns, we keep both values of the rest-frame time delay. As the luminosity state changes over time, the rest-frame time-delay adjusts accordingly, $\tau\propto L^{1/2}$. The resulting virial black hole mass remains consistent within the uncertainties since the line width behaves as $\Delta V\propto L^{-1/4}$. For NGC 4151, the virial black hole mass is $M_{\rm BH}=(4.14 \pm 0.73) \times 10^7\,M_{\odot}$ \citep{2006ApJ...647..901M}.}. The redshift range of this sample is $0.0033 \leq z \leq 1.89$, while the 3000 {\AA} luminosity of QSOs in the Mg II QSO-69 sample covers four orders of magnitude, $42.83 \leq \log_{10}{(L_{3000} [{\rm erg\,s^{-1}}])} \leq 46.79$. Both the low- and high-luminosity sources are beneficial for better determining the $R-L$ correlation relation. The Pearson correlation coefficient for the whole sample is $r=0.63$ with $p=5.60\times 10^{-9}$, while the Spearman correlation coefficient is $s=0.47$ with $p=4.52 \times 10^{-5}$, where $p$ expresses a two-sided $p$-value\footnote{The $p$-value relates to the hypothesis test, where the null hypothesis is that the two data sets, $\tau$ and $L_{3000}$, are uncorrelated. The $p$-value then estimates the probability with which these two uncorrelated data sets would yield the correlation coefficient that was inferred here.}. The RMS intrinsic scatter reaches $\sigma_{\rm ext} \sim 0.30$ dex for the standard $R-L$ relation, but it drops for the highly-accreting subsample, especially for extended versions of the $R-L$ relation \citep{Michal2021}. The sample is relatively homogeneous, with $\sim 83\%$ of the sources coming from the most recent SDSS-RM sample \citep{Homayouni2020} and $\sim 9\%$ of the sources from the previous SDSS-RM sample \citep{2016ApJ...818...30S}. This means that for most of the sources a consistent approach was used to infer the significant time-delay, mostly using the JAVELIN method that makes use of the damped random walk approach in fitting the continuum light curve \citep{2009ApJ...698..895K,2010ApJ...721.1014M,2010ApJ...708..927K,2011ApJ...735...80Z,2013ApJ...765..106Z,2016ApJ...819..122Z} as well as the CREAM that uses a random walk power spectral density prior of $P(f)\propto f^{-2}$ for the driving ionizing continuum \citep{2016MNRAS.456.1960S}. The remaining sources were analyzed typically by a combination of other methods, including a standard interpolation and discrete cross-correlation functions (ICCF and DCF, including the $z$-tranformed DCF), the $\chi^2$ method, and measures of data randomness/regularity \citep[see][ for overviews and applications to data]{czerny2013,2017ApJ...844..146C,2019AN....340..577Z,Michal2021}. The scatter along the RL correlation may be systematically increased due to the uncertainties of the time-delay analysis. For the largest SDSS-RM sample, \citet{Homayouni2020} analyzed the sample of 193 quasars in the redshift range of $0.35<z<1.7$, where they identified 57 significant time lags with the average false-positive rate of $11\%$. 24 sources out of them are further identified as a ``golden'' sample with the false-positive rate of $8\%$. In the older SDSS-RM sample of 6 quasars, the false-positive rate is comparable, at the level of $\sim 10\%-15\%$ for the reported significant lags \citep{2016ApJ...818...30S}. For the individual sources, a combination of more methods was typically employed to identify the consistent Mg II time delay, which was backed up by alias mitigation using bootstrap, pair-weighting, or Timmer-Koenig light-curve modelling, see e.g. \citet{Michal2021}.

\item[]{\bf Mg II QSO-9 sample}. This sample includes the last 9 QSOs listed in Table \ref{tab:MgQSOdata}. These data are from \citet{Zhefu2021}. They measured 9 significant Mg II lags using the first five years of data from the Dark Energy Survey \citep[DES, e.g.,][]{Flaugher2015} - Australian DES \citep[OzDES, e.g.,][]{Lidman2020} reverberation mapping project. The measurement sample spans the redshift range $\sim 1.1 - 1.7$. The lags are consistent with both the H$\beta$ $R-L$ relation determined by \citet{Bentz2013} and the Mg II $R-L$ relation of \citet{Homayouni2020}. For 9 Mg II time delays, the median false-positive rate is $4\%$.

\item[]{\bf Mg II QSO-78 sample.} This sample is the union of the Mg II QSO-69 and the Mg II QSO-9 samples. For the united sample, the Pearson correlation coefficient between $\tau$ and $L_{3000}$ is $r=0.63$ with $p=6.68\times 10^{-10}$ and the Spearman correlation coefficient is $s=0.50$ with $p=4.06\times 10^{-6}$, hence the correlation along the $R-L$ is slightly enhanced by adding MgII QSO-9 to the MgII QSO-69 sample. After the sample enlargement, the RMS scatter decreases only by $\sim 1.68\%$ from $\sim 0.30$ dex to $\sim 0.29$ dex. 
\end{itemize}

In this paper, we also use 31 $H(z)$ and 11 BAO measurements. The $H(z)$ data redshift range is $0.07 \leq z \leq 1.965$ and the BAO data redshift range is $0.0106 \leq z \leq 2.33$. The $H(z)$ data are given in Table 2 of \cite{Ryanetal2018} and the BAO data are listed in Table 1 of \cite{KhadkaRatra2021a}. Cosmological constraints obtained from the Mg II QSO samples are consistent with those obtained from the BAO + $H(z)$ data so we also jointly analyse the Mg II QSO-78 and BAO + $H(z)$ data sets.

\section{Methods}
\label{sec:methods}

\begin{figure}
    \includegraphics[width=\linewidth,right]{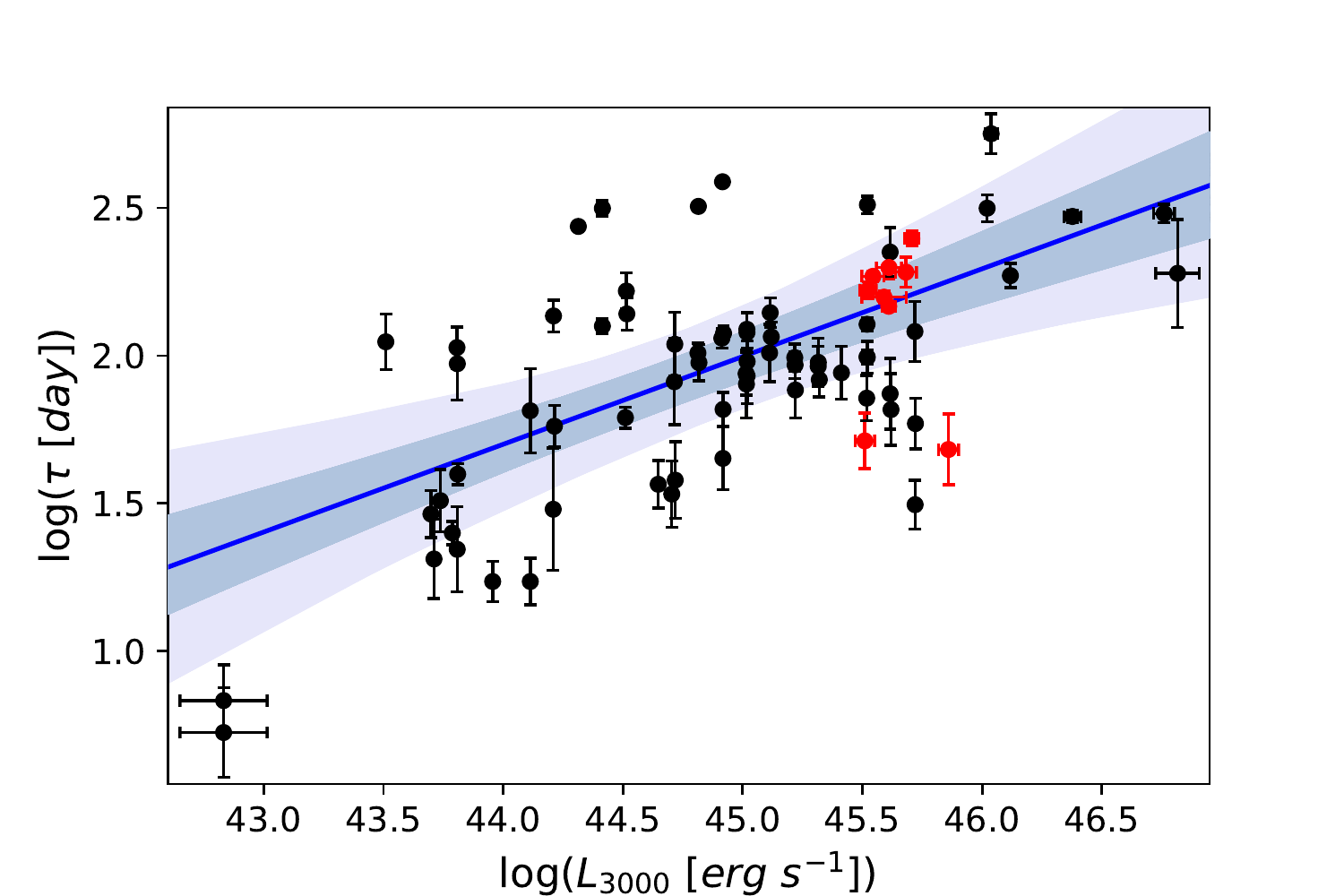}\par
\caption{$R-L$ correlation for 78 Mg II QSOs using the flat $\Lambda$CDM model. Black crosses show the Mg II QSO-69 sample and red crosses show the Mg II QSO-9 sample. Blue solid line is the $R-L$ correlation with best-fit parameter values for the QSO-78 data set. Blue and light gray shaded regions are the $1\sigma$ and $3\sigma$ confidence regions around the best-fit $R-L$ relation accounting only for the uncertainties in $\beta$ and $\gamma$.}
\label{fig:tau_L}
\end{figure}

The $R-L$ correlation relates the rest-frame time-delay of the Mg II broad line and the monochromatic luminosity of the QSO. For the sources used in this paper, this correlation can be seen in Fig.\ \ref{fig:tau_L}. The $R-L$ relation is usually expressed in the form
\begin{equation}
\label{eq:corr}
   \log \left({\frac{\tau} {\rm day}}\right) = \beta + \gamma \log\left({\frac{L_{3000}}{10^{44}\,{\rm erg\,s^{-1}}}}\right),
\end{equation}
where $\log$ = $\log_{10}$ and $L_{3000}$ and $\tau$ are the monochromatic luminosity of the quasar at 3000 {\AA} in the rest frame in units of erg s$^{-1}$ and the rest-frame time-delay of the Mg II line in units of day. Here $\beta$ and $\gamma$ are the correlation model free parameters and need to be determined from the data.   

The measured quantities are the time delay and the quasar flux. Expressing the luminosity in terms of the flux we obtain
\begin{equation}
\label{eq:corr_f}
   \log \left({\frac{\tau} {\rm day}}\right) = \beta + \gamma \log\left({\frac{F_{3000}}{10^{44}\,{\rm erg\,cm^{-2}\,s^{-1}}}}\right) + \gamma\log(4\pi) + 2\gamma\log\left(\frac{D_L}{\rm cm}\right),
\end{equation}
where $F_{3000}$ is the measured quasar flux at 3000 {\AA} in units of ${\rm erg\,cm^{-2}\,s^{-1}}$ and $D_L(z,p)$ is the luminosity distance in units of cm, which is a function of $z$ and the cosmological parameters $p$ of the cosmological model under study (see Sec.\ 2). The luminosity distance is
\begin{equation}
\label{eq:DM}
  \frac{H_0\sqrt{\left|\Omega_{k0}\right|}D_L(z, p)}{(1+z)} = 
    \begin{cases}
    {\rm sinh}\left[g(z)\right] & \text{if}\ \Omega_{k0} > 0, \\
    \vspace{1mm}
    g(z) & \text{if}\ \Omega_{k0} = 0,\\
    \vspace{1mm}
    {\rm sin}\left[g(z)\right] & \text{if}\ \Omega_{k0} < 0,
    \end{cases}   
\end{equation}
where
\begin{equation}
\label{eq:XCDM}
   g(z) = H_0\sqrt{\left|\Omega_{k0}\right|}\int^z_0 \frac{dz'}{H(z')},
\end{equation}
and $H(z)$ is the  Hubble parameter which is given in Sec.\ 2 for each cosmological model.

In a given cosmological model, eqs.\ (\ref{eq:corr_f}) and (\ref{eq:DM}) can be used to predict the rest-frame time-delay of the Mg II line for a quasar at known redshift. We can then compare the predicted and observed time-delays by using the likelihood function \citep{Dago2005}
\begin{equation}
\label{eq:chi2}
    \ln({\rm LF}) = -\frac{1}{2}\sum^{N}_{i = 1} \left[\frac{[\log(\tau^{\rm obs}_{X,i}) - \log(\tau^{\rm th}_{X,i})]^2}{s^2_i} + \ln(2\pi s^2_i)\right].
\end{equation}
Here $\ln$ = $\log_e$, $\tau^{\rm th}_{X,i}(p)$ and $\tau^{\rm obs}_{X,i}(p)$ are the predicted and observed time-delays at redshift $z_i$, and $s^2_i = \sigma^2_{\log{\tau_{\rm obs},i}} + \gamma^2 \sigma^2_{\log{F_{3000},i}} + \sigma_{\rm ext}^2$, where $\sigma_{\log{\tau_{\rm obs},i}}$ and $\sigma_{\log{F_{3000},i}}$ are the measurement error on the observed time-delay ($\tau^{\rm obs}_{X,i}(p)$) and the measured flux ($F_{3000}$) respectively, and $\sigma_{\rm ext}$ is the intrinsic dispersion of the $R-L$ relation.

QSO data alone cannot constrain $H_0$ because of the degeneracy between the correlation intercept parameter $\beta$ and $H_0$, so in this case we set $H_0$ to $70$ ${\rm km}\hspace{1mm}{\rm s}^{-1}{\rm Mpc}^{-1}$.

\begin{table}
	\centering
	\caption{Summary of the non-zero flat prior parameter ranges.}
	\label{tab:prior}
	\begin{threeparttable}
	\begin{tabular}{l|c}
	\hline
	Parameter & Prior range \\
	\hline
	$\Omega_bh^2$ & $[0, 1]$ \\
	$\Omega_ch^2$ & $[0, 1]$ \\
    $\Omega_{m0}$ & $[0, 1]$ \\
    $\Omega_{k0}$ & $[-2, 1]$ \\
    $\omega_{X}$ & $[-5, 0.33]$ \\
    $\alpha$ & $[0, 10]$ \\
    $\sigma_{\rm ext}$ & $[0, 5]$ \\
    $\beta$ & $[0, 10]$ \\
    $\gamma$ & $[0, 5]$ \\
	\hline
	\end{tabular}
    \end{threeparttable}
\end{table}

To determine cosmological model and $R-L$ parameter constraints from QSO-only data, we maximize the likelihood function given in eq.\ (\ref{eq:chi2}) and determine the best-fit values of all the free parameters and the corresponding uncertainties. The likelihood analysis for each data set and cosmological model is done using the Markov chain Monte Carlo (MCMC) method implemented in the \textsc{MontePython} code \citep{Brinckmann2019}. Convergence of the MCMC chains for each parameter is determined by using the Gelman-Rubin criterion $(R-1 < 0.05)$. For each free parameter we assume a top hat prior which is non-zero over the ranges given in Table \ref{tab:prior}.

To determine cosmological model parameter constraints from BAO + $H(z)$ data we use the method described in \cite{KhadkaRatra2021a}. To determine cosmological model and $R-L$ relation parameter constraints from QSO + BAO + $H(z)$ data we maximize the sum of the ln likelihood function given in eq.\ (\ref{eq:chi2}) and the BAO + $H(z)$ ln likelihood function given in eqs.\ (12) and (13) of \cite{KhadkaRatra2021a}.

For model comparisons, we compute the Akaike and Bayes Information Criterion ($AIC$ and $BIC$) values,
\begin{align}
\label{eq:AIC}
    AIC =& \chi^2_{\rm min} + 2d,\\
\label{eq:BIC}
    BIC =& \chi^2_{\rm min} + d\ln{N}\, ,
\end{align}
where $\chi^2_{\rm min} = -2 \ln({\rm LF}_{\rm max})$. Here $N$ is the number of data points, $d$ is the number of free parameters, and the degree of freedom $dof = N - d$. $AIC$ and $BIC$ penalize free parameters, while $\chi^2_{\rm min}$ does not, with $BIC$ more severely penalizing larger $d$ (than $AIC$ does) when $N \gtrsim 7.4$, as is the case for all data sets we consider here.  We also compute the differences, $\Delta AIC$ and $\Delta BIC$, with respect to the spatially-flat $\Lambda$CDM model $AIC$ and $BIC$ values. Positive $\Delta AIC$ or $\Delta BIC$ values indicate that the flat $\Lambda$CDM model is favored over the model under study. They provide weak, positive, and strong evidence for the flat $\Lambda$CDM model when they are in $[0, 2]$, $(2, 6]$, or $> 6$. Negative $\Delta AIC$ or $\Delta BIC$ values indicate that the model under study is favored over the flat $\Lambda$CDM model.

\begin{table*}
	\centering
	\small\addtolength{\tabcolsep}{-3.3pt}
	\small
	\caption{Unmarginalized one-dimensional best-fit parameters for Mg II QSO and BAO + $H(z)$ data sets. For each data set, $\Delta AIC$ and $\Delta BIC$ values are computed with respect to the $AIC$ and $BIC$  values of the flat  $\Lambda$CDM model.}
	\label{tab:BFP}
	\begin{threeparttable}
	\begin{tabular}{l|cccccccccccccccccccc} 
		\hline
		Model & Data set & $\Omega_{b}h^2$ & $\Omega_{c}h^2$& $\Omega_{\rm m0}$ & $\Omega_{\rm k0}$ & $\omega_{X}$ & $\alpha$ & $H_0$\tnote{a} & $\sigma_{\rm ext}$ & $\beta$ & $\gamma$ & $dof$ & $-2\ln({\rm LF}_{\rm max})$ & $AIC$ & $BIC$ & $\Delta AIC$ & $\Delta BIC$\\
		\hline
	    & Mg II QSO-69 & - & -& 0.155 & - & - & - & - & 0.288 & 1.667 & 0.290 & 65 & 29.56 & 37.56 & 46.50 & - & -\\
		Flat & Mg II QSO-78 & - & -& 0.138 & - & - & - &- & 0.281 & 1.666 & 0.283 & 74 & 30.16 & 38.16 & 47.58 & - & -\\
		$\Lambda$CDM & Mg II QSO-9 & - & - & 0.804 & - & - & - &- & $0.207$ & 2.154 & 0.002 & 5 & $-0.874$ & 7.126 & 7.91 & - & -\\
		&  B+H\tnote{b} & 0.024 & 0.119 & 0.298 & - & - & - &69.119&-&-&-& 39 & 23.66&29.66&34.87 & - & -\\
		& Q+B+H\tnote{c} & 0.024 & 0.119 & 0.300 & - & - & - &68.983&0.285&1.685&0.293& 115 & 53.96 & 63.96 & 77.90 & - & -\\
		\hline
		& Mg II QSO-69 & - & -& 0.357 & $-1.075$ & - &-&- & 0.274 & 1.612 & 0.364 & 64 & 23.50 & 33.50 & 44.67 & $-4.06$ & $-1.83$\\
		Non-flat & Mg II QSO-78 & - & - & 0.391 & $-$1.119 & - & - &- & 0.270 & 1.623 & 0.354 & 73 & 25.40 & 35.40 & 47.18 & $-2.76$ & $-0.40$\\
	    $\Lambda$CDM & Mg II QSO-9 &- & -& 0.664 & $-$0.759 & - & - &- & 0.211 & 2.157 & 0.001 & 4 & $-$0.88 & 9.12 & 10.11 & 2.00 & 2.20\\
	    & B+H\tnote{b} & 0.025 & 0.114 & 0.294 & 0.021 & - & - &68.701&-&-&-&38&23.60&31.60&38.55 & 1.94 & 3.68\\
	    & Q+B+H\tnote{c} & 0.024 & 0.117 & 0.298 & 0.012 & - & - &68.667&0.278&1.679&0.291&114&53.988&65.98&82.70 & 2.02 & 4.80\\
		\hline
		& Mg II QSO-69 &- & -& 0.003 & - & $-$4.998 &-&- & 0.277 & 1.353 & 0.233 & 64 & 23.98 & 33.98 & 45.15 & $-3.58$ & $-1.35$\\
		Flat & Mg II QSO-78 &- & -& 0.006 & - & $-$4.848 & - &- & 0.273 & 1.372 & 0.248 & 73 & 24.44 & 34.44 & 46.22 & $-3.72$ & $-1.36$\\
		XCDM & Mg II QSO-9 &- & -& 0.021 & - & $-$2.683 & - &- & 0.213 & 2.154 & 0.007 & 4 & $-$0.88 & 9.12 & 10.11 & 2.00 & 2.20\\
		& B+H\tnote{b} & 0.031 & 0.088 & 0.280 & - & $-$0.691 & - &65.036& - & - & -&38&19.66&27.66&34.61 & $-2.00$ & $-0.26$\\
		& Q+B+H\tnote{c} & 0.030 & 0.089 & 0.280 & - & $-$0.705 & - &65.097& 0.282 & 1.678 & 0.295&114&50.26&62.26&78.98 & $-1.70$ & 1.08\\
		\hline
		& Mg II QSO-69 &- & -& 0.043 & $-$0.091 & $-$2.727 &-&- & 0.262 & 1.455 & 0.293 & 63 & 17.96 & 29.96 & 43.36 & $-7.60$ & $-3.14$\\
		Non-flat & Mg II QSO-78 &- & - & 0.029 & $-$0.057 & $-$3.372 & - &- & 0.257 & 1.351 & 0.298 & 72 & 18.62 & 30.62 & 44.76 & $-7.54$ & $-2.82$\\
		XCDM & Mg II QSO-9 &- & -& 0.044 & 0.505 & $-$0.953 & - &- & 0.211 & 2.152 & 0.002 & 3 & $-$0.88 & 11.12 & 12.30 & 4.00 & 4.39\\
		& B+H\tnote{b} & 0.030 & 0.094 & 0.291 & $-$0.147 & $-$0.641 & - &65.204& - & - & -&37&18.34&28.34&37.03 & $-1.32$ & $2.16$\\
		& Q+B+H\tnote{c} & 0.029 & 0.100 & 0.295 & $-$0.159 & $-$0.643 & - &65.264& 0.292 & 1.682 & 0.298&113&48.94&62.94& 82.45 & $-1.02$ & 4.55\\
		\hline
		& Mg II QSO-69 &- & -& 0.149 & - & - &9.150&- & 0.288 & 1.668 & 0.286 & 64 & 29.56 & 39.56 & 50.73 & 2.00 & 4.23\\
		Flat & Mg II QSO-78 &- & -& 0.171 & - & - & 8.777 &- & 0.281 & 1.672 & 0.285 & 73 & 30.16 & 40.16 & 51.94 & 2.00 & 4.36\\
		$\phi$CDM & Mg II QSO-9 &- & -& 0.377 & - & - & 7.795 &- & 0.208 & 2.148 & 0.003 & 4 & $-$0.88 & 9.12 & 10.11 & 2.00 & 2.20\\
		& B+H\tnote{b} & 0.033 & 0.080 & 0.265 & - & - & 1.445 &65.272& - & - & -&38&19.56&27.56&34.51 & $-2.10$ & $-0.36$\\
		& Q+B+H\tnote{c} & 0.031 & 0.086 & 0.272 & - & - & 1.212 &65.628&0.280&1.693&0.288& 114 & 50.12&62.12&78.84 & $-1.84$ & 0.94\\
		\hline
		& Mg II QSO-69 &- & -& 0.439 & $-$0.440 & - & 9.540 &- & 0.287 & 1.672 & 0.307 & 63 & 29.18 & 41.18 & 54.58 & 3.62 & 8.08\\
		Non-flat & Mg II QSO-78 &- & - & 0.341 & $-$0.333 & - & 5.637 & - & 0.282 & 1.671 & 0.296 & 72 & 29.76 & 41.76 & 55.90 & 3.60 & 8.32\\
		$\phi$CDM & Mg II QSO-9 &- & -& 0.879 & $-$0.185 & - & 7.644 &- & 0.212 & 2.155 & 0.001 & 3 & $-0.88$ & 11.12 & 12.30 & 4.00 & 4.39\\
		& B+H\tnote{b} & 0.035 & 0.078 & 0.261 & $-$0.155 & - & 2.042 &65.720& - & - & -&37&18.16&28.16&36.85 & $-1.50$ & 1.98\\
		& Q+B+H\tnote{c} & 0.033 & 0.082 & 0.265 & $-$0.160 & - & 1.902 &65.876& 0.284 & 1.682 & 0.297&113&48.72&62.72&82.23 & $-1.24$ & 4.33\\
		\hline
	\end{tabular}
	\begin{tablenotes}
    \item[a]${\rm km}\hspace{1mm}{\rm s}^{-1}{\rm Mpc}^{-1}$. $H_0$ is set to $70$ ${\rm km}\hspace{1mm}{\rm s}^{-1}{\rm Mpc}^{-1}$ for the QSO-only data analyses.
    \item[b]${\rm BAO}+H(z)$.
    \item[c]Mg II QSO-78 + ${\rm BAO}+H(z)$.
    \end{tablenotes}
    \end{threeparttable}
\end{table*}

\begin{sidewaystable*}
\begin{adjustbox}{angle=180}
\centering
\small\addtolength{\tabcolsep}{0.0pt}
\begin{threeparttable}
\caption{Marginalized one-dimensional best-fit parameters with 1$\sigma$ confidence intervals, or 1$\sigma$ or 2$\sigma$ limits, for the Mg II QSO and BAO + $H(z)$ data sets.}
\label{tab:1d_BFP2}
\setlength{\tabcolsep}{1.3mm}{
\begin{tabular}{lccccccccccccc}
\hline
Model & Data & $\Omega_{b}h^2$ & $\Omega_{c}h^2$& $\om$ & $\ol$\tnote{a} & $\ok$ & $\omega_{X}$ & $\alpha$ & $H_0$\tnote{b} & $\sigma_{\rm ext}$ & $\beta$ & $\gamma$ \\
\hline
Flat \lcdm\ & Mg II QSO-69 &-&-& $0.240^{+0.450}_{-0.170}$ & $0.758^{+0.172}_{-0.448}$ & - & - & - &-& $0.301^{+0.024}_{-0.032}$ & $1.699^{-0.059}_{-0.059}$ & $0.300^{+0.049}_{-0.049}$\\
& Mg II QSO-78 &-&-& $0.270^{+0.400}_{-0.210}$ & $0.729^{+0.211}_{-0.399}$ & - & - & - &-& $0.292^{+0.022}_{-0.029}$ & $1.700^{-0.058}_{-0.058}$ & $0.297^{+0.046}_{-0.046}$\\
& Mg II QSO-9 &-&-& $> 0.088$ & $< 0.912$ & - & - & - &-& $0.257^{+0.113}_{-0.073}$ & $1.712^{-0.368}_{-0.732}$ & $0.296^{+0.414}_{-0.296}$\\
& BAO+H\tnote{c}& $0.024^{+0.003}_{-0.003}$ & $0.119^{+0.008}_{-0.008}$ & $0.299^{+0.015}_{-0.017}$ & - & - & - & - &$69.300^{+1.800}_{-1.800}$&-&-&-\\
& Q+B+H\tnote{d} & $0.024^{+0.003}_{-0.003}$ & $0.119^{+0.007}_{-0.008}$ & $0.299^{+0.015}_{-0.017}$ & - & - & - & - &$69.300^{+1.800}_{-1.800}$& $0.291^{+0.022}_{-0.029}$&$1.682^{+0.054}_{-0.054}$&$0.293^{+0.043}_{-0.043}$\\
\hline
Non-flat \lcdm\ & Mg II QSO-69 &-&-& $0.681^{+0.219}_{-0.301}$ & $1.785^{+0.335}_{-0.985}$ & $-1.296^{+0.926}_{-0.684}$ & - &-&-& $0.297^{+0.025}_{-0.032}$ & $1.674^{+0.065}_{-0.065}$ & $0.324^{+0.052}_{-0.060}$\\
& Mg II QSO-78 &-&-& $0.726^{+0.153}_{-0.397}$ & $1.712^{+0.298}_{-1.122}$ & $-1.169^{+1.269}_{-0.511}$ & - &-&-& $0.289^{+0.023}_{-0.029}$ & $1.680^{+0.063}_{-0.063}$ & $0.317^{+0.048}_{-0.055}$\\
& Mg II QSO-9 &-&-& $> 0.126$ & $0.661^{+0.639}_{-0.660}$ & $> -1.51$ & - &-&-& $0.256^{+0.112}_{-0.076}$ & $1.678^{+0.412}_{-0.668}$ & $0.317^{+0.433}_{-0.277}$\\
& BAO+H\tnote{c}& $0.025^{+0.004}_{-0.004}$ & $0.113^{+0.019}_{-0.019}$ & $0.292^{+0.023}_{-0.023}$ & $0.667^{+0.093}_{+0.081}$ & $-0.014^{+0.075}_{-0.075}$ & - & - &$68.700^{+2.300}_{-2.300}$&-&-&-\\
& Q+B+H\tnote{d} & $0.025^{+0.004}_{-0.005}$ & $0.115^{+0.018}_{-0.018}$ & $0.294^{+0.023}_{-0.023}$ & $0.675^{+0.092}_{+0.079}$ & $0.031^{+0.094}_{-0.110}$ & - & - &$68.800^{+2.200}_{-2.200}$&$0.292^{+0.022}_{-0.029}$&$1.681^{+0.055}_{-0.055}$&$0.293^{+0.044}_{-0.044}$\\
\hline
Flat XCDM & Mg II QSO-69 &-&-& (< 0.496, 1$\sigma$) & - & - & $< -0.393$ & - &-& $0.298^{+0.025}_{-0.032}$ & $1.675^{+0.085}_{-0.109}$ & $0.297^{+0.049}_{-0.049}$\\
& Mg II QSO-78 &-&-& (< 0.500, 1$\sigma$) & - & - & $< -0.367$ & - &-& $0.291^{+0.024}_{-0.030}$ & $1.640^{+0.120}_{-0.074}$ & $0.294^{+0.046}_{-0.046}$\\
& Mg II QSO-9 &-&-& --- & - & - & --- & - &-& $0.261^{+0.113}_{-0.082}$ & $1.614^{+0.476}_{-0.624}$ & $0.294^{+0.046}_{-0.046}$\\
& BAO+H\tnote{c} & $0.030^{+0.005}_{-0.005}$ & $0.093^{+0.019}_{-0.017}$ & $0.282^{+0.021}_{-0.021}$ & - & - & $-0.744^{+0.140}_{-0.097}$ & - &$65.800^{+2.200}_{-2.500}$& - & - & -\\
&Q+B+H\tnote{d}& $0.030^{+0.004}_{-0.005}$ & $0.093^{+0.019}_{-0.016}$ & $0.283^{+0.023}_{-0.020}$ & - & - & $-0.750^{+0.150}_{-0.100}$ & - &$65.800^{+2.200}_{-2.600}$& $0.292^{+0.022}_{-0.029}$ & $1.680^{+0.055}_{-0.055}$ & $0.294^{+0.044}_{-0.044}$\\
\hline
Non-flat XCDM & Mg II QSO-69&-&- & $0.287^{+0.513}_{-0.087}$ & - & $-0.339^{+0.559}_{-0.681}$ & $-1.138^{+0.738}_{-2.362}$ & - &-& $0.297^{+0.025}_{-0.032}$ & $1.672^{+0.088}_{-0.107}$ & $0.318^{+0.051}_{-0.057}$\\
& Mg II QSO-78 &-&-& $0.373^{+0.407}_{-0.133}$ & - & $-0.303^{+0.523}_{-0.677}$ & $< 0.246$ & - &-& $0.289^{+0.023}_{-0.029}$ & $1.640^{+0.120}_{-0.079}$ & $0.314^{+0.048}_{-0.053}$\\
& Mg II QSO-9 &-&-& --- & - & $0.000^{+0.810}_{-0.540}$ & $- 0.728^{+0.788}_{-2.262}$ & - &-& $0.256^{+0.111}_{-0.077}$ & $1.802^{+0.318}_{-0.702}$ & $0.197^{+0.493}_{-0.177}$\\
& BAO+H\tnote{c} & $0.029^{+0.005}_{-0.005}$ & $0.099^{+0.021}_{-0.021}$ & $0.293^{+0.027}_{-0.027}$ & - & $-0.120^{+0.130}_{-0.130}$ & $-0.693^{+0.130}_{-0.077}$ & - &$65.900^{+2.400}_{-2.400}$& - & - & -\\
& Q+B+H\tnote{d} & $0.029^{+0.005}_{-0.006}$ & $0.099^{+0.021}_{-0.021}$ & $0.293^{+0.027}_{-0.027}$ & - & $-0.120^{+0.130}_{-0.130}$ & $-0.700^{+0.140}_{-0.079}$ & - &$66.000^{+2.200}_{-2.500}$& $0.292^{+0.022}_{-0.029}$ & $1.682^{+0.055}_{-0.055}$ & $0.296^{+0.044}_{-0.044}$\\
\hline
Flat $\phi$CDM & Mg II QSO-69 &-&-& $0.264^{+0.406}_{-0.214}$ & - & - & - & --- &-& $0.301^{+0.025}_{-0.033}$ & $1.697^{+0.063}_{-0.057}$ & $0.299^{+0.049}_{-0.049}$\\
& Mg II QSO-78 &-&-& $0.276^{+0.394}_{-0.216}$ & - & - & - & --- &-& $0.293^{+0.022}_{-0.029}$ & $1.699^{+0.061}_{-0.055}$ & $0.296^{+0.046}_{-0.046}$\\
& Mg II QSO-9 &-&-& --- & - & - & - & --- &-& $0.247^{+0.106}_{-0.067}$ & $1.831^{+0.279}_{-0.651}$ & $0.167^{+0.443}_{-0.147}$\\
& BAO+H\tnote{c} & $0.032^{+0.006}_{-0.003}$ & $0.081^{+0.017}_{-0.017}$ & $0.266^{+0.023}_{-0.023}$ & - & - & - & $1.530^{+0.620}_{-0.850}$ &$65.100^{+2.100}_{-2.100}$& - & - & -\\
& Q+B+H\tnote{d} & $0.032^{+0.006}_{-0.003}$ & $0.081^{+0.018}_{-0.018}$ & $0.266^{+0.024}_{-0.024}$ & - & - & - & $1.510^{+0.620}_{-0.890}$ &$65.200^{+2.100}_{-2.100}$& $0.292^{+0.022}_{-0.029}$ & $1.680^{+0.055}_{-0.055}$ & $0.295^{+0.044}_{-0.044}$\\
\hline
Non-flat $\phi$CDM & Mg II QSO-69 &-&-& --- & - & $-0.009^{+0.399}_{-0.361}$ & - & --- &-& $0.301^{+0.025}_{-0.033}$ & $1.700^{+0.058}_{-0.058}$ & $0.301^{+0.049}_{-0.049}$\\
& Mg II QSO-78 &-&-& --- & - & $-0.011^{+0.401}_{-0.359}$ & - & --- &-& $0.292^{+0.022}_{-0.029}$ & $1.702^{+0.055}_{-0.055}$ & $0.298^{+0.045}_{-0.045}$\\
& Mg II QSO-9 &-&-& --- & - & $0.000^{+0.430}_{-0.330}$ & - & --- &-& $0.254^{+0.113}_{-0.074}$ & $1.793^{+0.317}_{-0.793}$ & $0.214^{+0.516}_{-0.194}$\\
& BAO+H\tnote{c} & $0.032^{+0.006}_{-0.004}$ & $0.085^{+0.017}_{-0.021}$ & $0.271^{+0.024}_{-0.028}$ & - & $-0.080^{+0.100}_{-0.100}$ & - & $1.660^{+0.670}_{-0.830}$ &$65.500^{+2.500}_{-2.500}$& - & - & -\\
& Q+B+H\tnote{d} & $0.032^{+0.007}_{-0.004}$ & $0.086^{+0.018}_{-0.022}$ & $0.272^{+0.024}_{-0.029}$ & - & $-0.090^{+0.100}_{-0.120}$ & - & $1.660^{+0.670}_{-0.850}$ &$65.600^{+2.200}_{-2.200}$& $0.292^{+0.022}_{-0.029}$ & $1.681^{+0.055}_{-0.055}$ & $0.295^{+0.044}_{-0.044}$\\
\hline
\end{tabular}}
\begin{tablenotes}
\item[a]In our analyses $\Omega_{\Lambda}$ is a derived parameter and in each case $\Omega_{\Lambda}$ chains are derived using the current energy budget equation $\Omega_{\Lambda}= 1-\Omega_{m0}-\Omega_{k0}$ (where $\Omega_{k0}=0$ in the flat $\Lambda$CDM model). From these chains, using the \textsc{python} package \textsc{getdist} \citep{Lewis_2019}, we determine best-fit values and uncertainties for $\Omega_{\Lambda}$. We also use this \textsc{python} package to plot the likelihoods and compute the best-fit values and uncertainties of the free parameters.
\item[b]${\rm km}\hspace{1mm}{\rm s}^{-1}{\rm Mpc}^{-1}$. $H_0$ is set to $70$ ${\rm km}\hspace{1mm}{\rm s}^{-1}{\rm Mpc}^{-1}$ for the QSO-only data analyses.
\item[c]BAO + $H(z)$.
\item[d]Mg II QSO-78 + BAO + $H(z)$.
\end{tablenotes}
\end{threeparttable}
\end{adjustbox}
\end{sidewaystable*}

\section{Results}
\label{sec:QSO}
\subsection{Mg II QSO-69, Mg II QSO-9, and Mg II QSO-78 data constraints}
\label{MgIIresults}

Results for the Mg II QSO-69, QSO-9, and QSO-78 data sets are given in Tables \ref{tab:BFP} and \ref{tab:1d_BFP2}. The unmarginalized best-fit parameter values are listed in Table \ref{tab:BFP} and the marginalized one-dimensional best-fit parameter values and limits are given in Table \ref{tab:1d_BFP2}. The one-dimensional likelihood distributions and the two-dimensional likelihood contours for the Mg II QSO-69 and Mg II QSO-78 data sets are shown in blue and olive, respectively, in Figs.\ 2--4 and corresponding plots for the Mg II QSO-9 data set are shown in blue in Figs.\ 5--7.

\begin{figure*}
\begin{multicols}{2}
    \includegraphics[width=\linewidth]{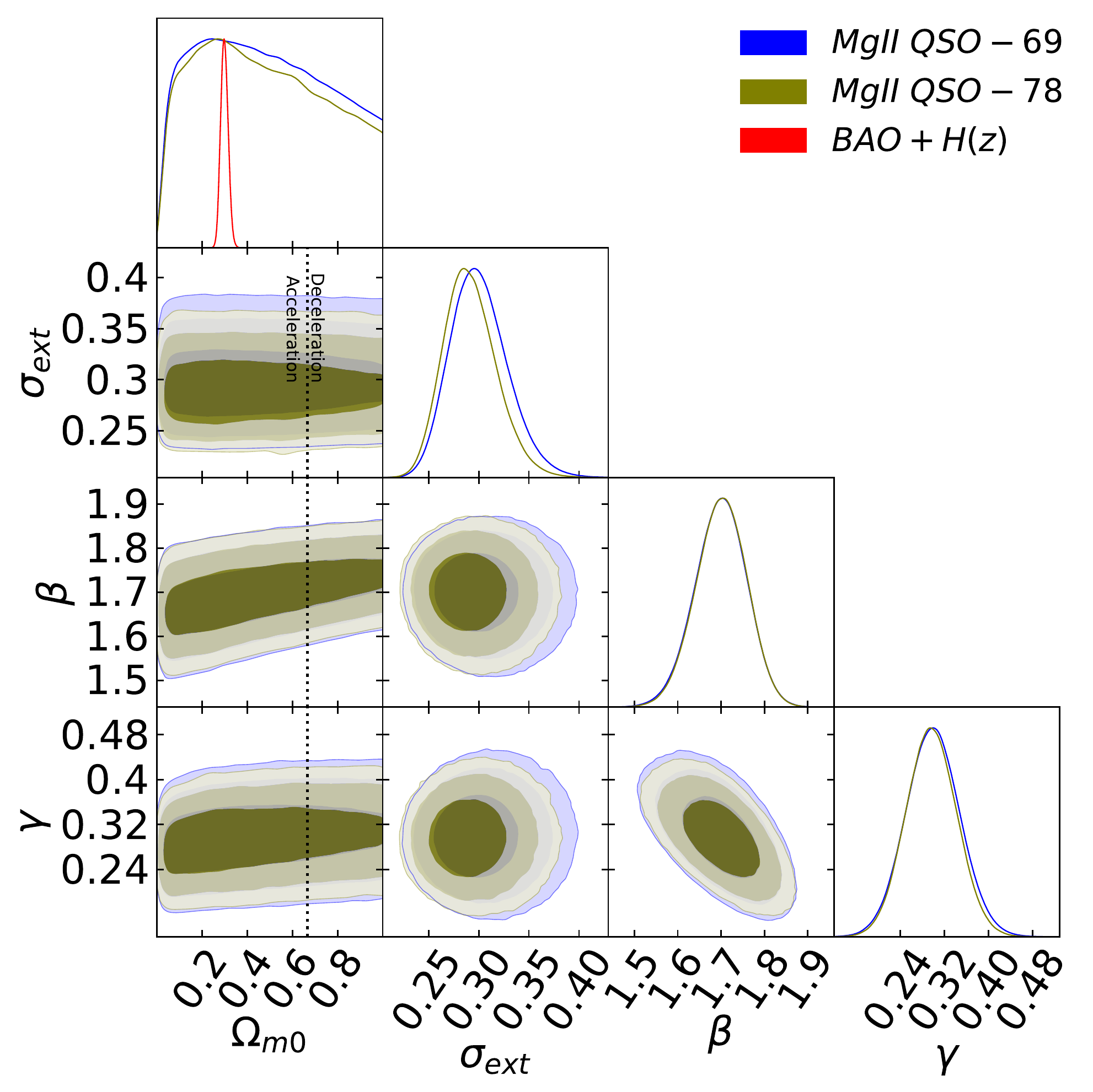}\par
    \includegraphics[width=\linewidth]{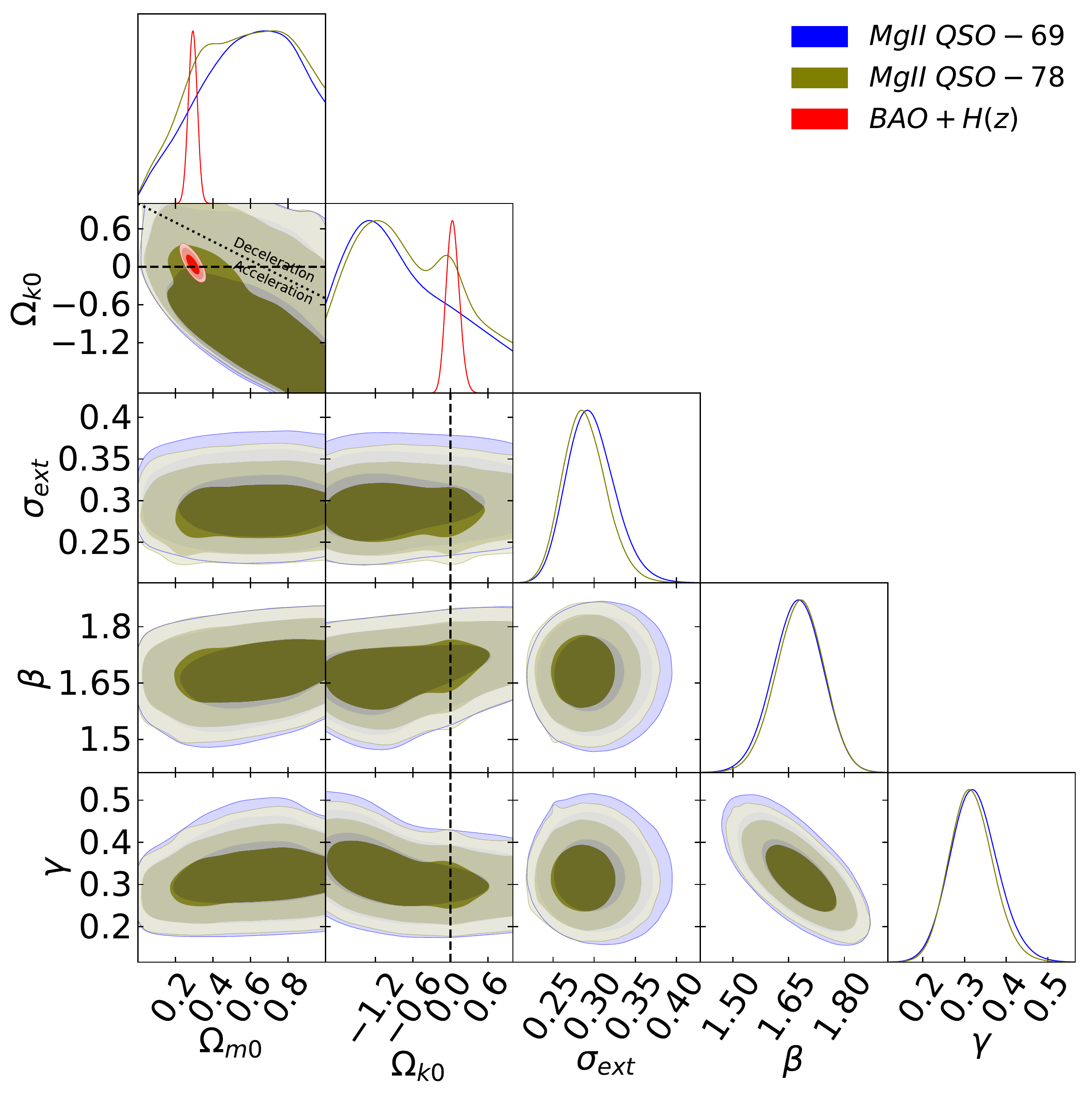}\par
\end{multicols}
\caption{One-dimensional likelihood distributions and two-dimensional likelihood contours at 1$\sigma$, 2$\sigma$, and 3$\sigma$ confidence levels using Mg II QSO-69 (blue), Mg II QSO-78 (olive), and BAO + $H(z)$ (red) data for all free parameters. Left panel shows the flat $\Lambda$CDM model. The black dotted vertical lines are the zero acceleration lines with currently accelerated cosmological expansion occurring to the left of the lines. Right panel shows the non-flat $\Lambda$CDM model. The black dotted sloping line in the $\Omega_{k0}-\Omega_{m0}$ subpanel is the zero acceleration line with currently accelerated cosmological expansion occurring to the lower left of the line. The black dashed horizontal or vertical line in the $\Omega_{k0}$ subpanels correspond to $\Omega_{k0} = 0$.}
\label{fig:Eiso-Ep}
\end{figure*}

\begin{figure*}
\begin{multicols}{2}
    \includegraphics[width=\linewidth]{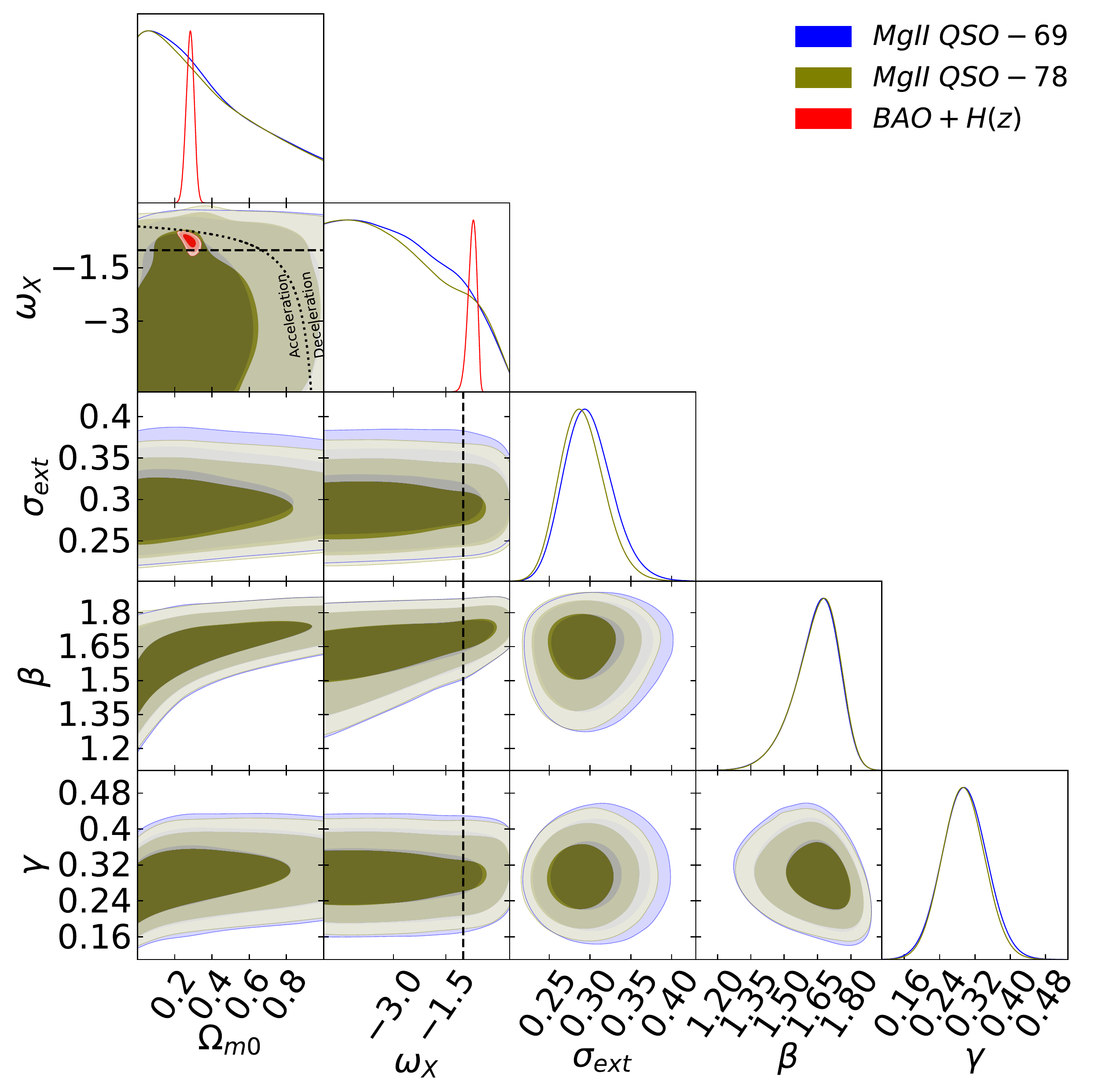}\par
    \includegraphics[width=\linewidth]{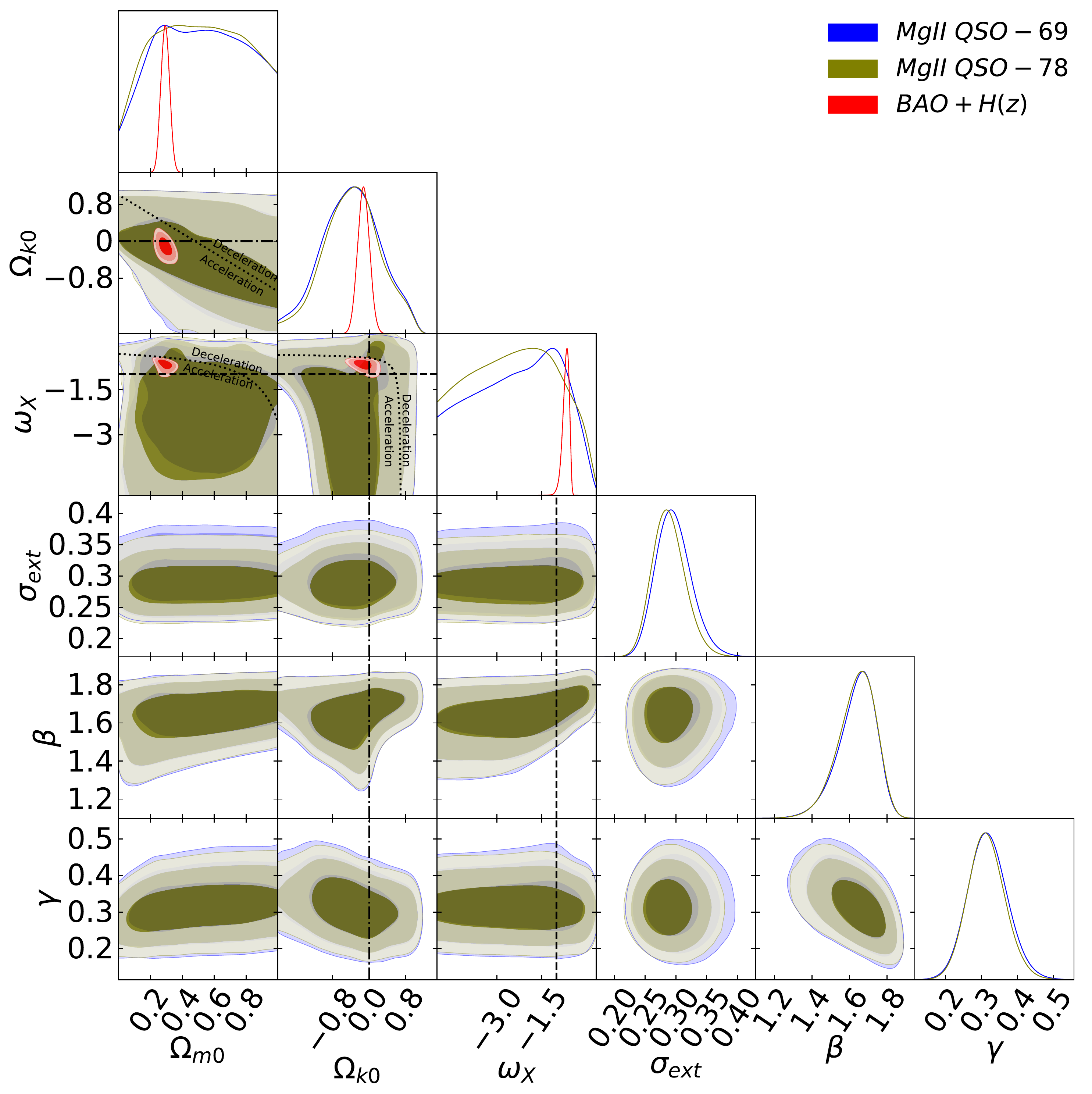}\par
\end{multicols}
\caption{One-dimensional likelihood distributions and two-dimensional likelihood contours at 1$\sigma$, 2$\sigma$, and 3$\sigma$ confidence levels using Mg II QSO-69 (blue), Mg II QSO-78 (olive), and BAO + $H(z)$ (red) data for all free parameters. Left panel shows the flat XCDM parametrization. The black dotted curved line in the $\omega_X-\Omega_{m0}$ subpanel is the zero acceleration line with currently accelerated cosmological expansion occurring below the line and the black dashed straight lines correspond to the $\omega_X = -1$ $\Lambda$CDM model. Right panel shows the non-flat XCDM parametrization. The black dotted lines in the $\Omega_{k0}-\Omega_{m0}$, $\omega_X-\Omega_{m0}$, and $\omega_X-\Omega_{k0}$ subpanels are the zero acceleration lines with currently accelerated cosmological expansion occurring below the lines. Each of the three lines is computed with the third parameter set to the BAO + $H(z)$ data best-fit value given in Table 3. The black dashed straight lines correspond to the $\omega_X = -1$ $\Lambda$CDM model. The black dotted-dashed straight lines correspond to $\Omega_{k0} = 0$.}
\label{fig:Eiso-Ep}
\end{figure*}

\begin{figure*}
\begin{multicols}{2}
    \includegraphics[width=\linewidth]{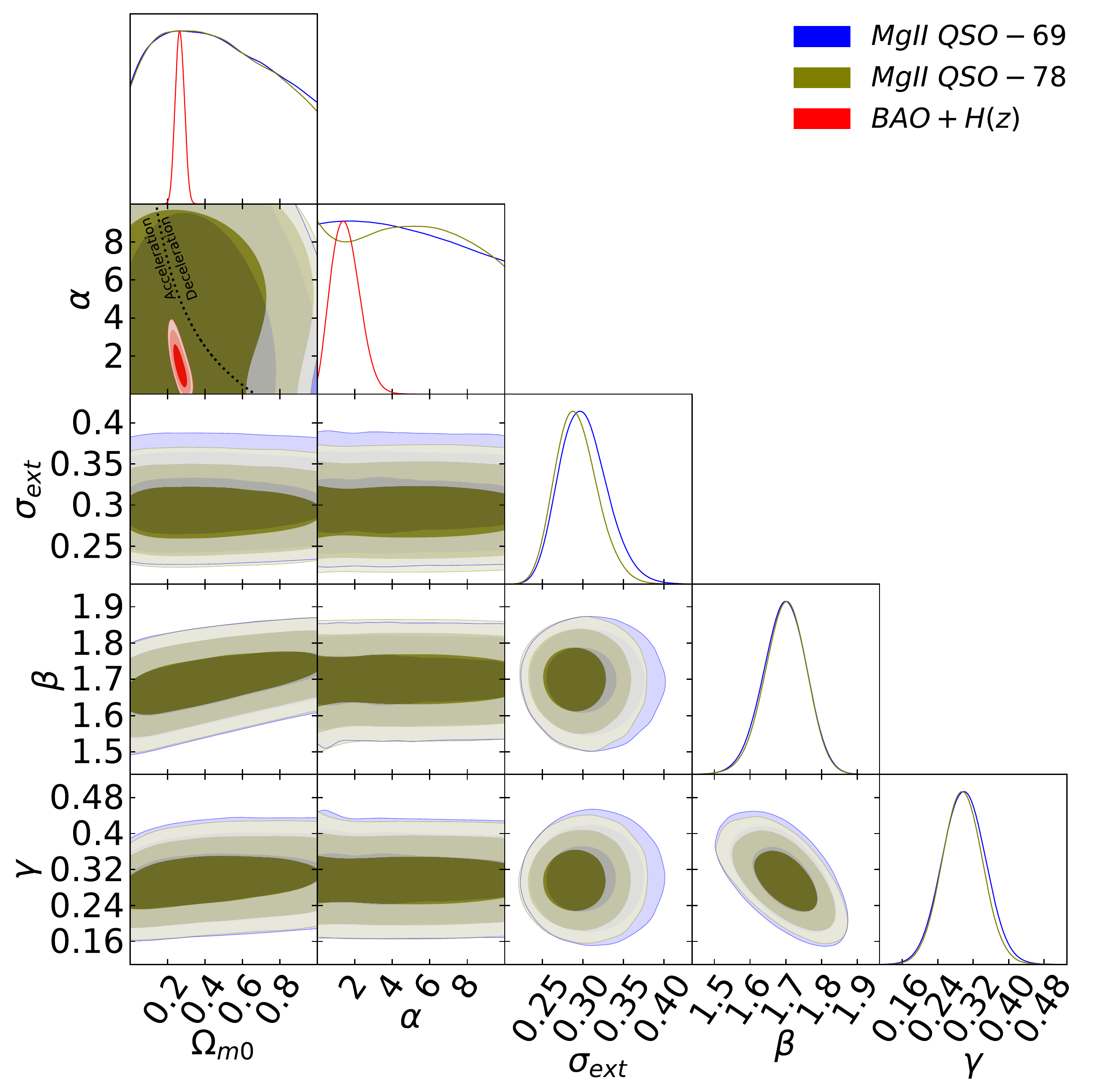}\par
    \includegraphics[width=\linewidth]{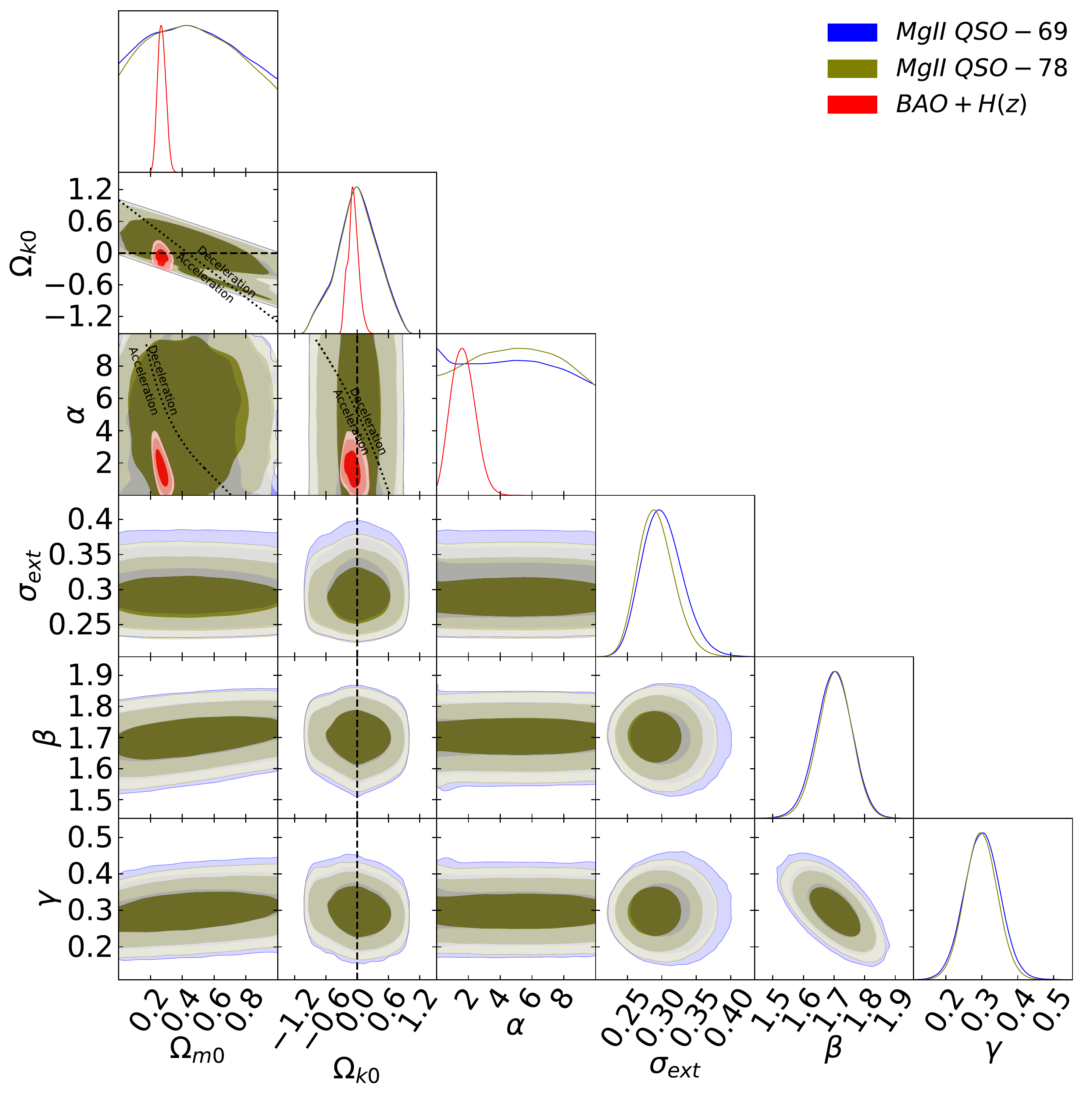}\par
\end{multicols}
\caption{One-dimensional likelihood distributions and two-dimensional likelihood contours at 1$\sigma$, 2$\sigma$, and 3$\sigma$ confidence levels using Mg II QSO-69 (blue), Mg II QSO-78 (olive), and BAO + $H(z)$ (red) data for all free parameters. The $\alpha = 0$ axes correspond to the $\Lambda$CDM model. Left panel shows the flat $\phi$CDM model. The black dotted curved line in the $\alpha - \Omega_{m0}$ subpanel is the zero acceleration line with currently accelerated cosmological expansion occurring to the left of the line. Right panel shows the non-flat $\phi$CDM model. The black dotted lines in the $\Omega_{k0}-\Omega_{m0}$, $\alpha-\Omega_{m0}$, and $\alpha-\Omega_{k0}$ subpanels are the zero acceleration lines with currently accelerated cosmological expansion occurring below the lines. Each of the three lines is computed with the third parameter set to the BAO + $H(z)$ data best-fit value given in Table 3. The black dashed straight lines correspond to $\Omega_{k0} = 0$.}
\label{fig:Eiso-Ep}
\end{figure*}

\begin{figure*}
\begin{multicols}{2}
    \includegraphics[width=\linewidth]{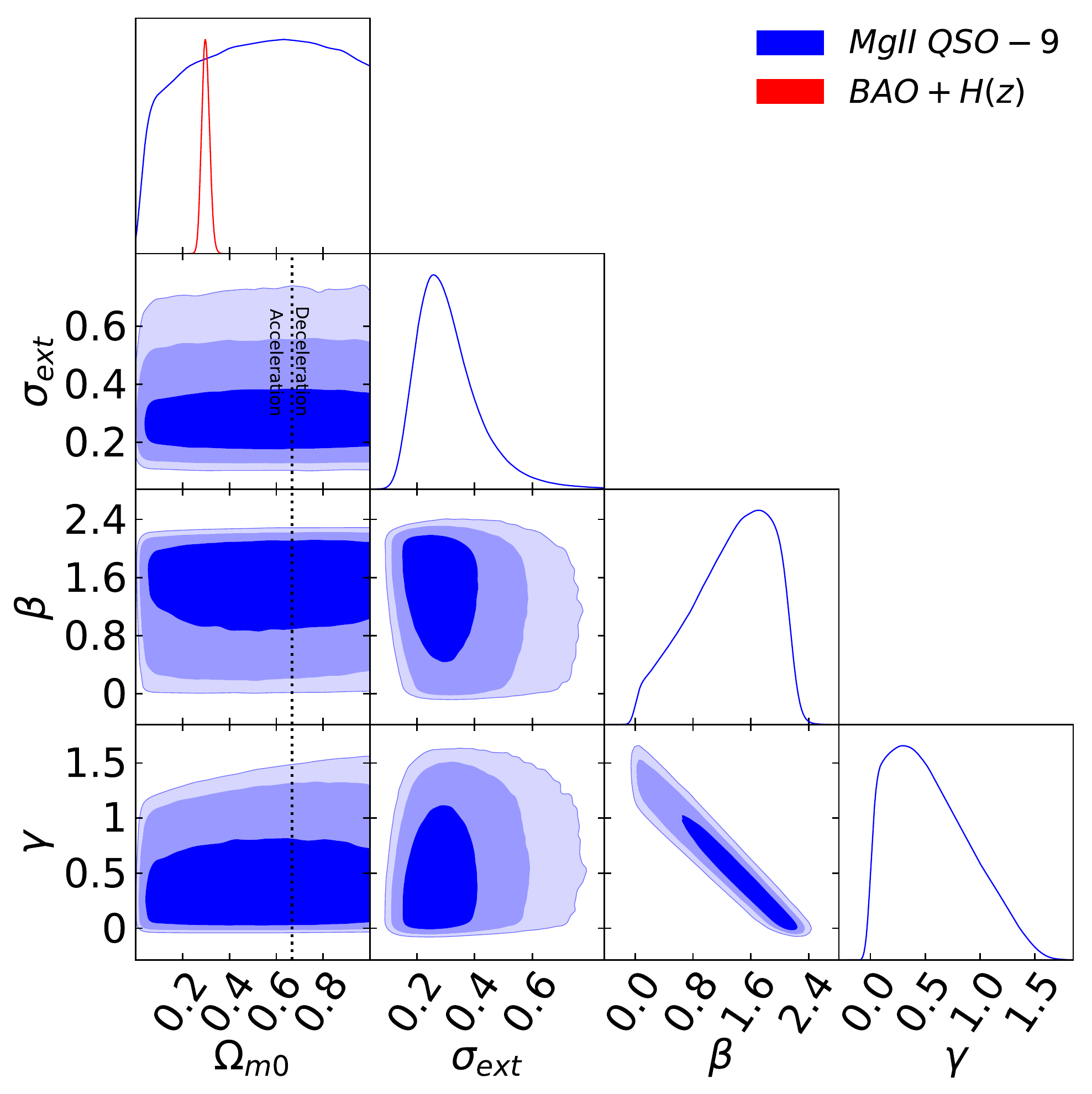}\par
    \includegraphics[width=\linewidth]{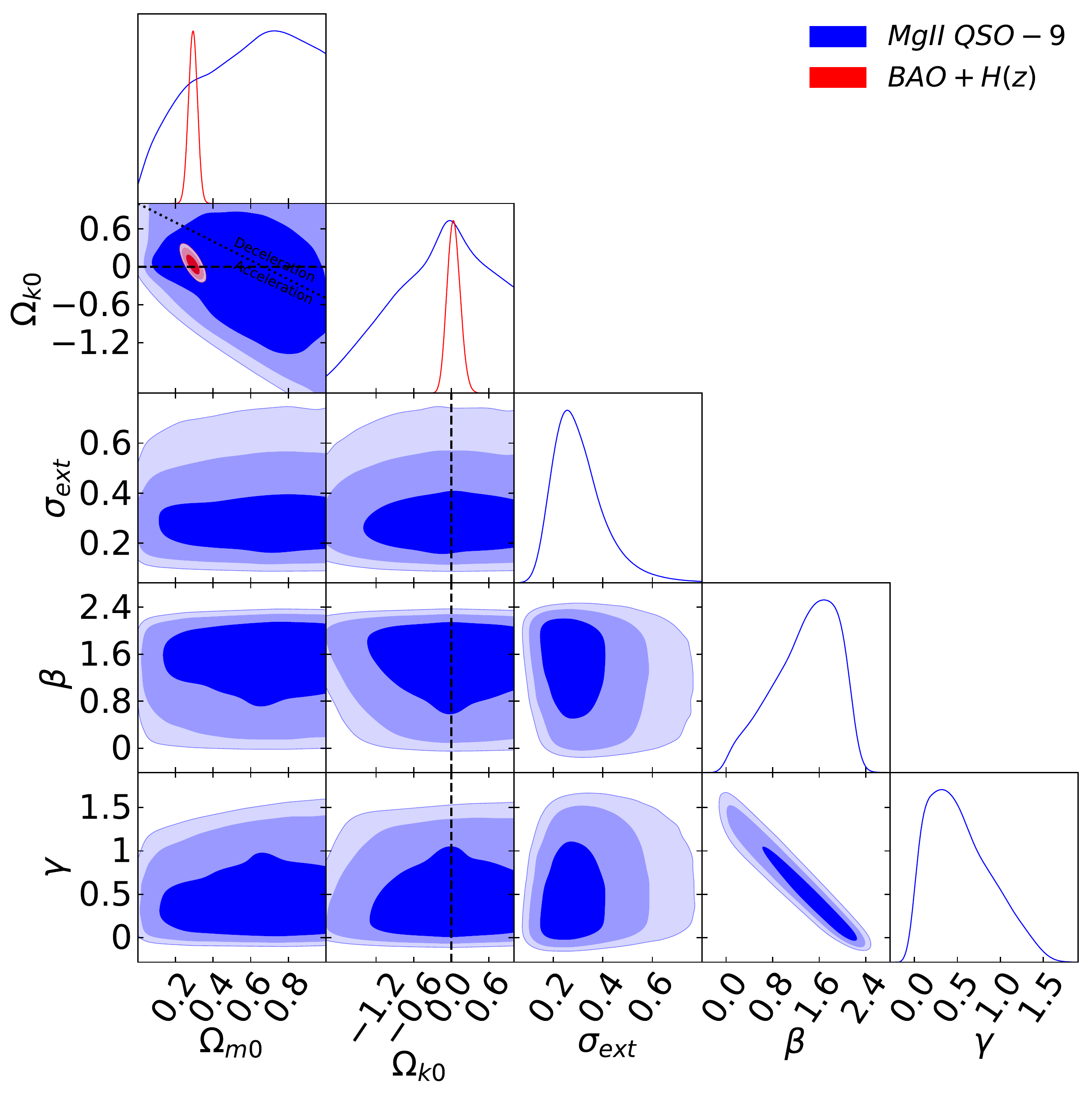}\par
\end{multicols}
\caption{One-dimensional likelihood distributions and two-dimensional likelihood contours at 1$\sigma$, 2$\sigma$, and 3$\sigma$ confidence levels using Mg II QSO-9 (blue), and BAO + $H(z)$ (red) data for all free parameters. Left panel shows the flat $\Lambda$CDM model. The black dotted vertical lines are the zero acceleration lines with currently accelerated cosmological expansion occurring to the left of the lines. Right panel shows the non-flat $\Lambda$CDM model. The black dotted sloping line in the $\Omega_{k0}-\Omega_{m0}$ subpanel is the zero acceleration line with currently accelerated cosmological expansion occurring to the lower left of the line. The black dashed horizontal or vertical line in the $\Omega_{k0}$ subpanels correspond to $\Omega_{k0} = 0$.}
\label{fig:Eiso-Ep}
\end{figure*}

\begin{figure*}
\begin{multicols}{2}
    \includegraphics[width=\linewidth]{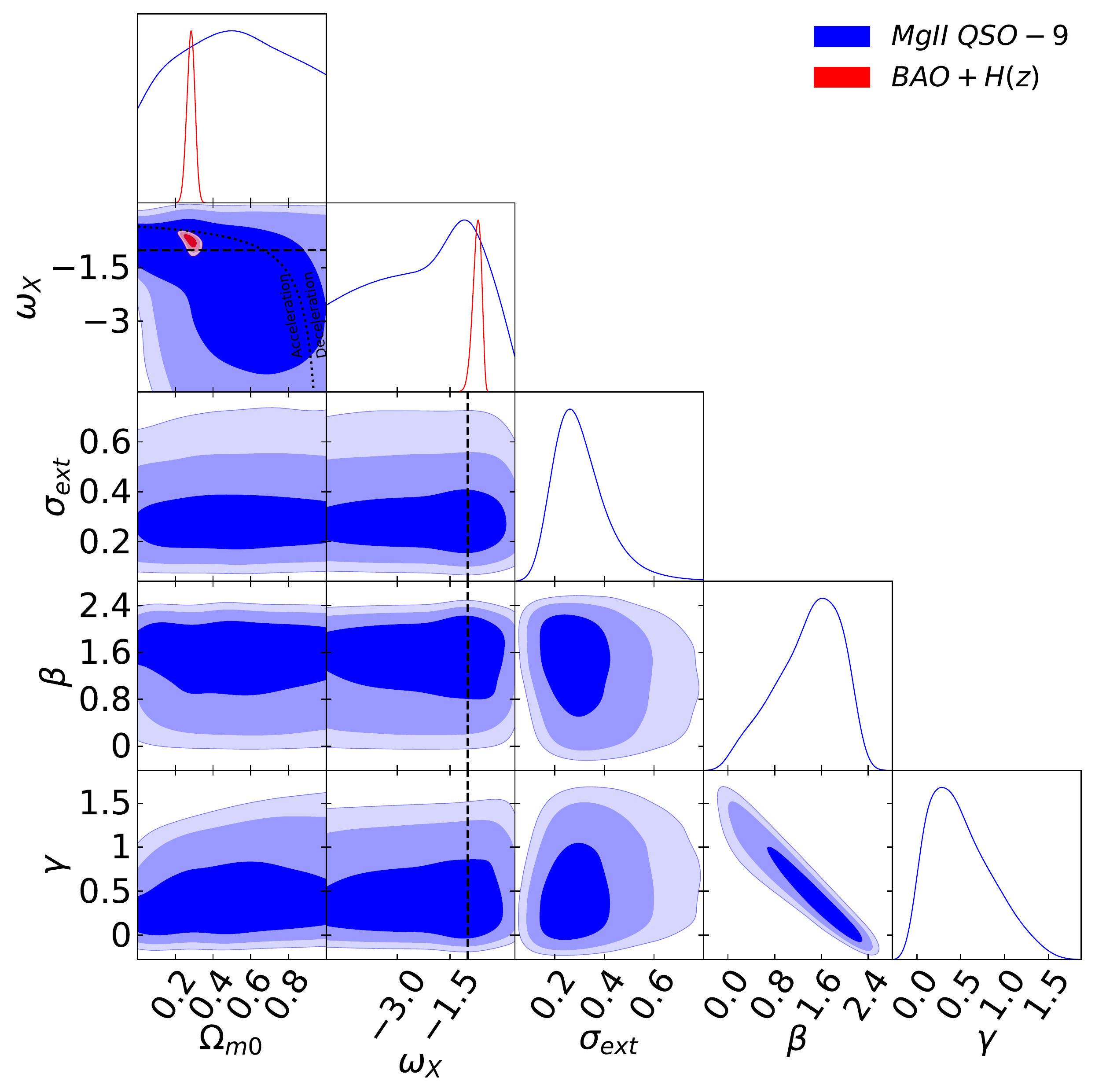}\par
    \includegraphics[width=\linewidth]{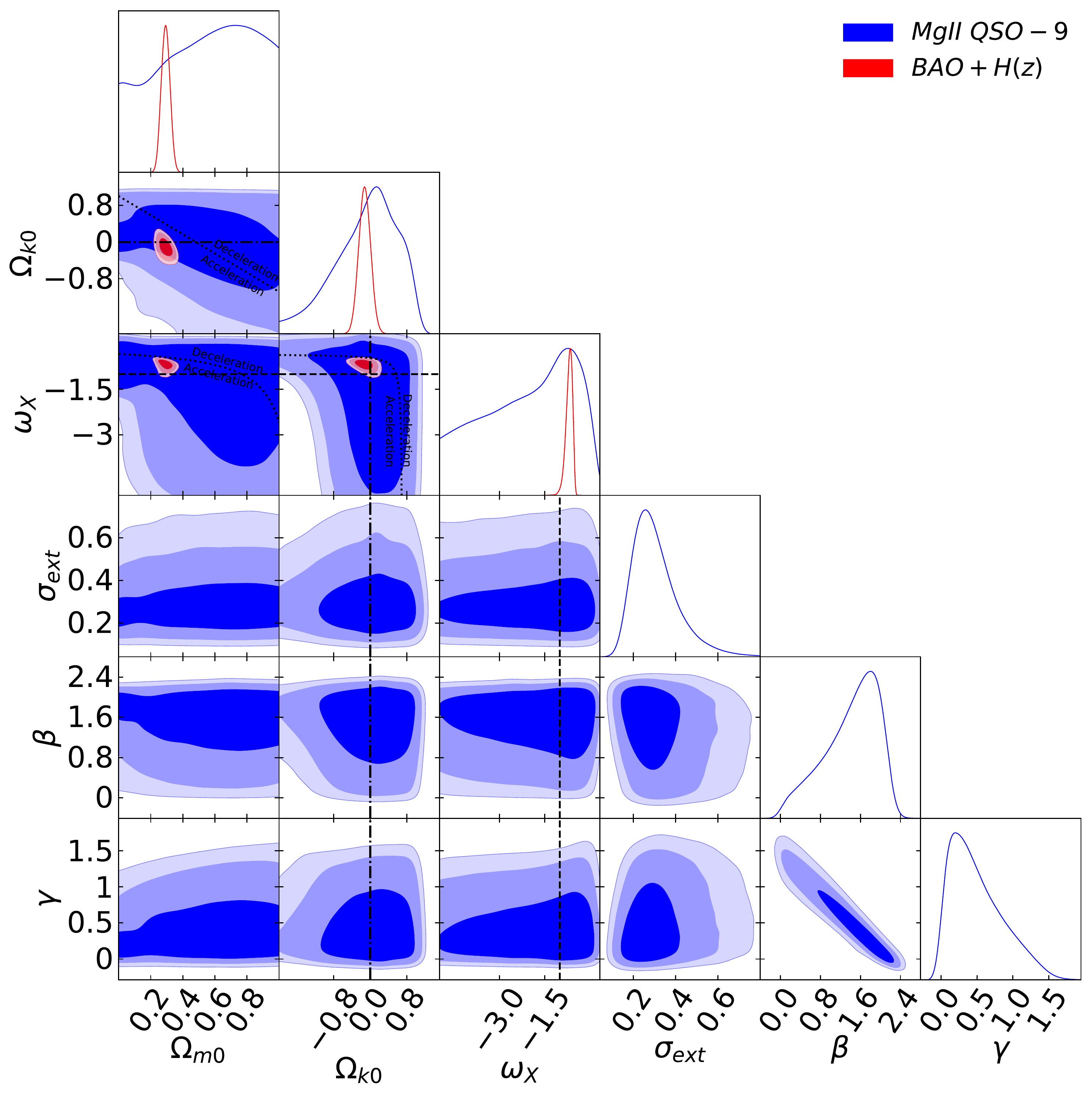}\par
\end{multicols}
\caption{One-dimensional likelihood distributions and two-dimensional likelihood contours at 1$\sigma$, 2$\sigma$, and 3$\sigma$ confidence levels using Mg II QSO-9 (blue), and BAO + $H(z)$ (red) data for all free parameters. Left panel shows the flat XCDM parametrization. The black dotted curved line in the $\omega_X-\Omega_{m0}$ subpanel is the zero acceleration line with currently accelerated cosmological expansion occurring below the line and the black dashed straight lines correspond to the $\omega_X = -1$ $\Lambda$CDM model. Right panel shows the non-flat XCDM parametrization. The black dotted lines in the $\Omega_{k0}-\Omega_{m0}$, $\omega_X-\Omega_{m0}$, and $\omega_X-\Omega_{k0}$ subpanels are the zero acceleration lines with currently accelerated cosmological expansion occurring below the lines. Each of the three lines is computed with the third parameter set to the BAO + $H(z)$ data best-fit value given in Table 3. The black dashed straight lines correspond to the $\omega_X = -1$ $\Lambda$CDM model. The black dotted-dashed straight lines correspond to $\Omega_{k0} = 0$.}
\label{fig:Eiso-Ep}
\end{figure*}

\begin{figure*}
\begin{multicols}{2}
    \includegraphics[width=\linewidth]{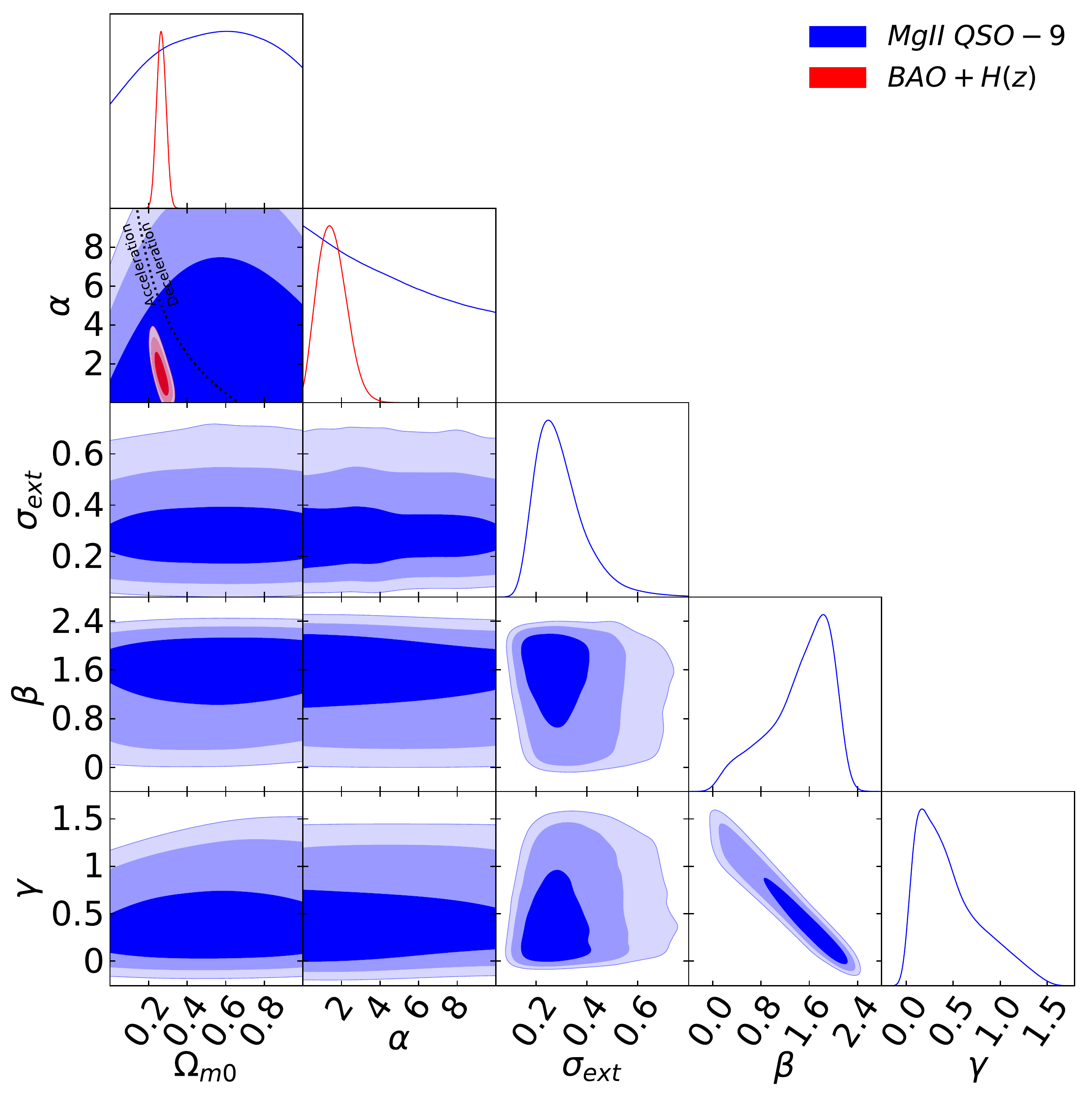}\par
    \includegraphics[width=\linewidth]{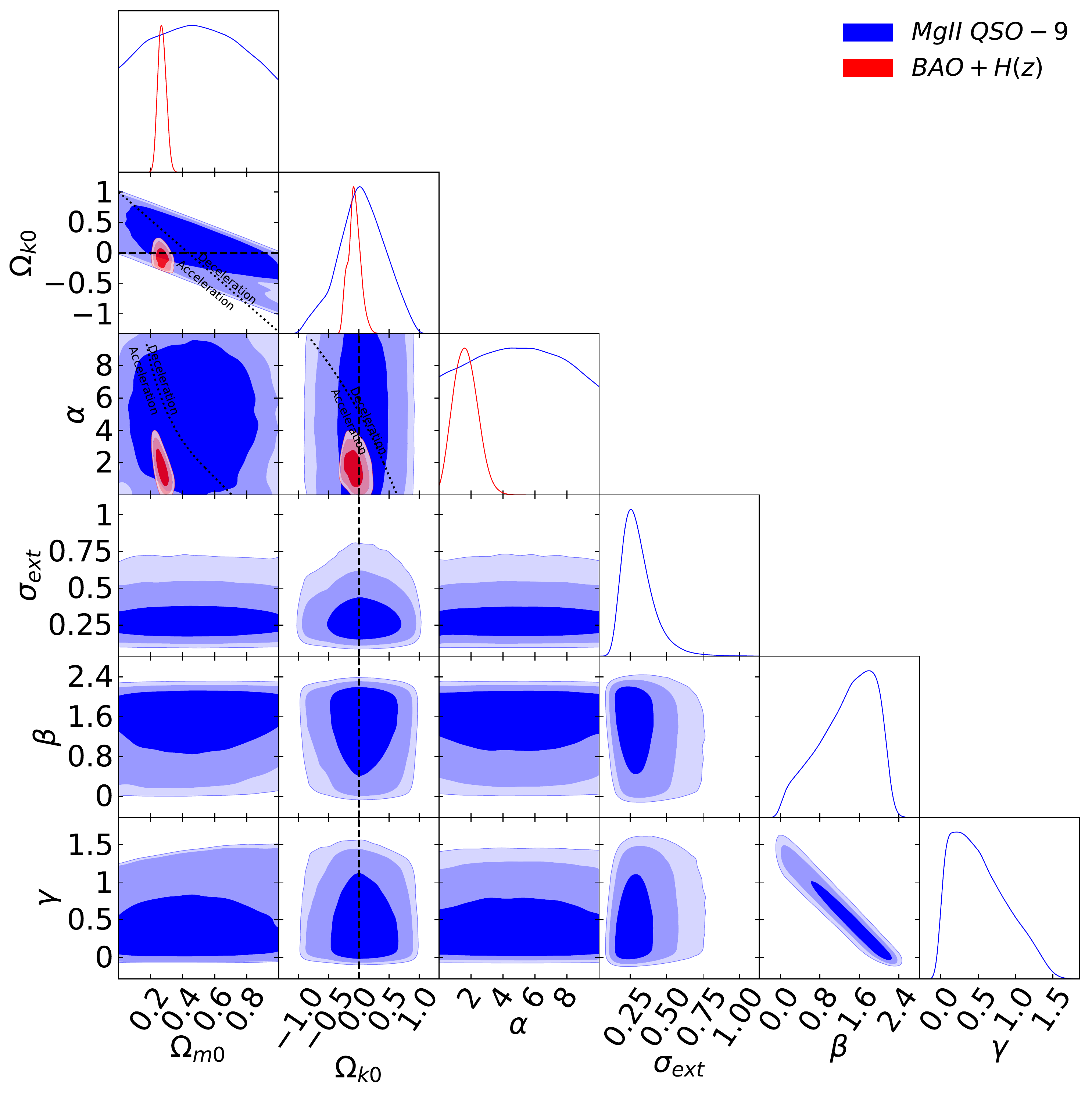}\par
\end{multicols}
\caption{One-dimensional likelihood distributions and two-dimensional likelihood contours at 1$\sigma$, 2$\sigma$, and 3$\sigma$ confidence levels using Mg II QSO-9 (blue), and BAO + $H(z)$ (red) data for all free parameters. The $\alpha = 0$ axes correspond to the $\Lambda$CDM model. Left panel shows the flat $\phi$CDM model. The black dotted curved line in the $\alpha - \Omega_{m0}$ subpanel is the zero acceleration line with currently accelerated cosmological expansion occurring to the left of the line. Right panel shows the non-flat $\phi$CDM model. The black dotted lines in the $\Omega_{k0}-\Omega_{m0}$, $\alpha-\Omega_{m0}$, and $\alpha-\Omega_{k0}$ subpanels are the zero acceleration lines with currently accelerated cosmological expansion occurring below the lines. Each of the three lines is computed with the third parameter set to the BAO + $H(z)$ data best-fit value given in Table 3. The black dashed straight lines correspond to $\Omega_{k0} = 0$.}
\label{fig:Eiso-Ep}
\end{figure*}

The Mg II QSO-9 data set is small and so constraints derived using these data have larger error bars than those determined from the QSO-69 data. From Table \ref{tab:1d_BFP2} and Figs.\ 2--7, we see that the QSO-9 and QSO-69 constraints are consistent and so it is reasonable to use the combined QSO-78 data to constrain parameters. 

From Table \ref{tab:1d_BFP2} we see that the $R-L$ relation parameters $\beta$ and $\gamma$ for each data set, QSO-9, QSO-69, and QSO-78, have values that are independent of the cosmological model assumed in the analysis. This validates the basic assumption of the $R-L$ relation and means that these sources can be used as standardizable candles to constrain cosmological model parameters. For these three data sets, the best-fit values of $\beta$ are $\sim 1.7$ and the best-fit values of $\gamma$ are $\sim 0.3$. The Mg II $R-L$ relation is thus shallower than the value predicted by the simple photoionization model ($\gamma = 0.5$). This is not a problem from a photoionization point of view because it appears that the broad Mg II line is emitted towards the outer part of the BLR and it exhibits a weaker response to the continuum variation than do the Balmer emission lines \citep{guo2020}; see however \citet{Michal2020} for a significant correlation coefficient of $\sim 0.8$ and the presence of the intrinsic Baldwin effect for the luminous quasar HE 0413-4031. In addition, the Mg II line is a resonance line that is mostly collisionally excited, while Balmer lines are recombination lines. This can qualitatively affect the slope of the $R-L$ relation for the Mg II line in comparison with the Balmer lines. However, \citet{Mary2020} and \citet{Michal2021} found that by separating the sample into low and high accretors, it is possible to recover the expected value in both cases, i.e. the slope increases from $\sim 0.3$. This result supports the existence of the $R-L$ correlation for Mg II QSOs, which is also consistent with the theoretical findings of \citet{guo2020}, who predict the existence of the global Mg II  $R-L$ correlation, while the weaker response of Mg II to the continuum variations can affect the $R-L$ correlation slope for some individual sources, but apparently not all, or the epochs of correlated line light curve may be interrupted by a decorrelated light curve \citep[BLR ``holidays''; see also the study of NGC 5548;][for an example]{2019ApJ...882L..30D}. Given that there is a significant Mg II QSO $R-L$ correlation, as long as there are no significant unaccounted-for errors, an $R-L$ relation slope $\sim 0.3$ (instead of $\sim 0.5$) does not invalidate the cosmological usage of Mg II QSOs. Another free parameter of the $R-L$ relation is the intrinsic dispersion ($\sigma_{\rm ext}$). The minimum value of $\sigma_{\rm ext}$, $\sim 0.25$ dex, is obtained using the Mg II QSO-9 data set and the maximum value of $\sigma_{\rm ext}$, $\sim 0.3$ dex, is obtained using the Mg II QSO-69 data set.

For the combined Mg II QSO-78 data, $\sigma_{\rm ext} \sim 0.29$ dex. This is smaller than the $\sigma_{\rm ext} \sim 0.39$ dex for the best available gamma-ray burst data set of 118 standardizable-candle GRBs spanning $0.3399 \leq z \leq 8.2$ \citep{Khadkaetal2021} and a little larger than the $\sigma_{\rm ext} \sim 0.24$ dex for the best available QSO X-ray and UV flux data set of 1019 standardizable-candle QSOs spanning $0.009 \leq z \leq 1.479$ \citep{KhadkaRatra2021a}.

The scatter $\sigma_{\rm ext}$ appears to be driven by the accretion-rate as shown by \citet{Michal2020} and \citet{Michal2021}. In principle, the scatter could partially be mitigated by adding an independent observational quantity to the RL relation correlated with the accretion rate, see \citet{Mary2020} for the analysis using the relative Fe II strength or fractional AGN variability parameters. This would, however, add one more nuisance parameter besides $\beta$ and $\gamma$ in the fitting scheme, and the overall effect on constraining cosmological parameters needs to be studied in detail in a future study. Furthermore, a homogeneous time-delay analysis applied to all the sources may also help to mitigate a fraction of the scatter, especially for a larger sample, since some sources exhibit more comparable peaks in correlation space, see e.g. \citet{2019ApJ...880...46C}, which creates a systematic uncertainty in the time-delay determination.

From Figs.\ 2--4 we see that for the Mg II QSO-78 data set the likelihoods favor the part of cosmological model parameter space that is consistent with currently-accelerating cosmological expansion, with the non-flat $\phi$CDM model being somewhat of an outlier.  

\begin{figure}
    \includegraphics[width=\linewidth,right]{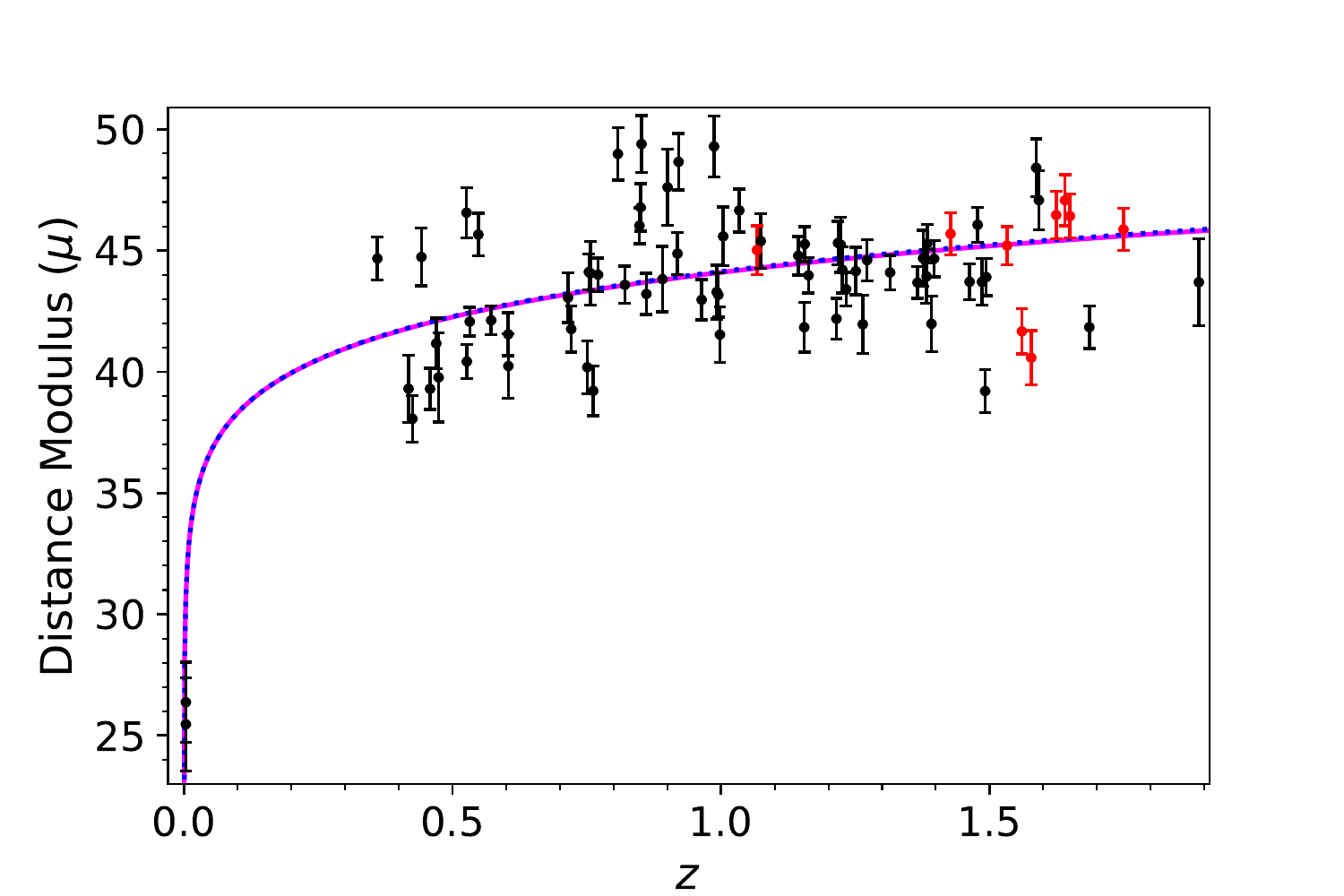}\par
\caption{Hubble diagram of 78 Mg II QSOs in the best-fit flat $\Lambda$CDM model. Magenta solid line is the prediction for the best-fit flat $\Lambda$CDM model with $\Omega_{m0}=0.27$ from the Mg II QSO-78 data set. Black and red data points are the observed distance moduli and corresponding uncertainties for the Mg II QSO-69 and Mg II QSO-9 samples respectively in the best-fit QSO-78 flat $\Lambda$CDM model. The blue dotted line shows the standard flat $\Lambda$CDM model with $\Omega_{m0}=0.3$.}
\label{fig:Hubble_diagram}
\end{figure}

From Table \ref{tab:1d_BFP2}, for the Mg II QSO-69 data set, the minimum value of $\Omega_{m0}$, $0.240^{+0.450}_{-0.170}$, is obtained in the spatially-flat $\Lambda$CDM model and the maximum value of $\Omega_{m0}$, $0.681^{+0.219}_{-0.301}$, is in the spatially non-flat $\Lambda$CDM model. These data cannot constrain $\Omega_{m0}$ in the flat XCDM parametrization or the non-flat $\phi$CDM model. For the Mg II QSO-9 data, the value of $\Omega_{m0}$ is determined to be > 0.088 and > 0.126, at 2$\sigma$, in the flat and non-flat $\Lambda$CDM model respectively. These data cannot constrain $\Omega_{m0}$ in the four other models. For the Mg II QSO-78 data, the minimum value of $\Omega_{m0}$, $0.270^{+0.400}_{-0.210}$, is in the flat $\Lambda$CDM model and the maximum value of $\Omega_{m0}$, $0.726^{+0.153}_{-0.397}$, is in the non-flat $\Lambda$CDM model. These data cannot constrain $\Omega_{m0}$ in the flat XCDM parametrization or the non-flat $\phi$CDM model. All $\Omega_{m0}$ values obtained using these QSO data sets are consistent with those from BAO + $H(z)$ data or other well-established cosmological probes such as CMB anisotropy or Type Ia supernova measurements. In Fig.\ \ref{fig:Hubble_diagram} we plot the Hubble diagram of the 78 Mg II QSOs and this figure shows that this QSO Hubble diagram is consistent with that of a flat $\Lambda$CDM model with $\Omega_{m0} = 0.3$.

From Table \ref{tab:1d_BFP2} and Figs.\ 2--4, we see that currently-available Mg II QSO data set at most only weak constraints on $\Omega_{\Lambda}$, $\Omega_{k0}$, $\omega_X$, and $\alpha$.\footnote{In the spatially non-flat $\phi$CDM model, $\Omega_{\phi}(z, \alpha)$ is obtained from the numerical solutions of the equations of motion and its current value always lies in the range $0 \leq \Omega_{\phi}(0, \alpha) \leq 1$. This restriction on $\Omega_{\phi}(0,\alpha)$ can be seen in the non-flat $\phi$CDM model plots in Figs.\ 4 and 7 in the form of straight-line contour boundaries in the $\Omega_{m0}-\Omega_{k0}$ subpanels.}

Table \ref{tab:BFP} lists, for all three QSO data sets, the values of $AIC$, $BIC$, and their differences, $\Delta AIC$ and $\Delta BIC$, with respect to the $AIC$ and $BIC$ values for the spatially-flat $\Lambda$CDM model. From the $AIC$ and $BIC$ values, for the Mg II QSO-69 and Mg II QSO-78 data sets, the most favored case is the non-flat XCDM parametrization while non-flat $\phi$CDM is least favored. From the $AIC$ and $BIC$ values, for the Mg II QSO-9 data set, the most favored case is the flat $\Lambda$CDM model while the non-flat XCDM parametrization and the $\phi$CDM model are least favored. From the $\Delta AIC$ values, only in the non-flat XCDM parametrization do the Mg II QSO-69 and Mg II QSO-78 data sets provide strong evidence against the spatially-flat $\Lambda$CDM model. From the $\Delta BIC$ values, the Mg II QSO-69 and Mg II QSO-78 data sets provide strong evidence against only the non-flat $\phi$CDM model.

\subsection{BAO + $H(z)$ and Mg II QSO-78 + BAO + $H(z)$ data constraints}
\label{com_con}

\begin{figure*}
\begin{multicols}{2}
    \includegraphics[width=\linewidth]{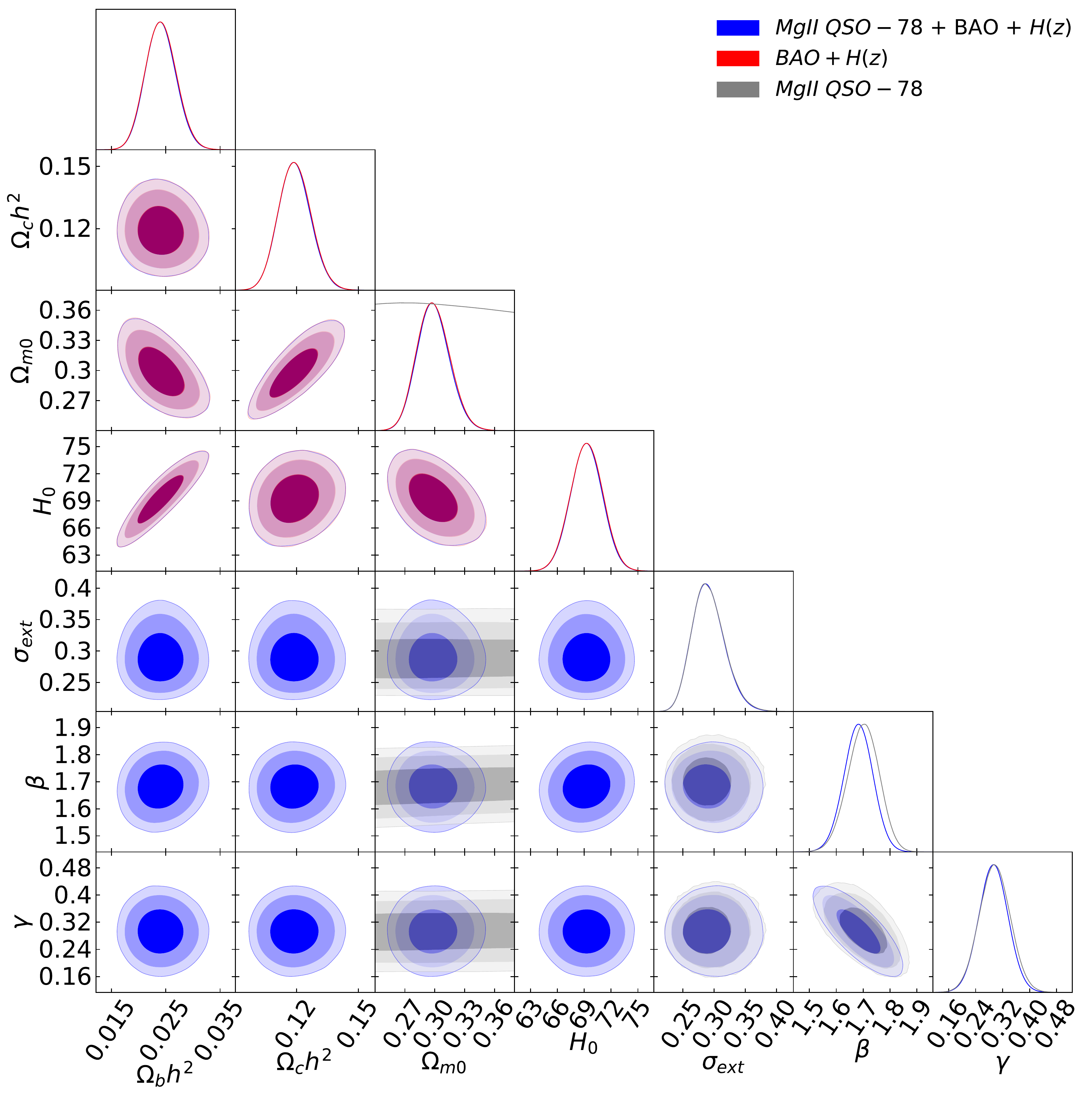}\par
    \includegraphics[width=\linewidth]{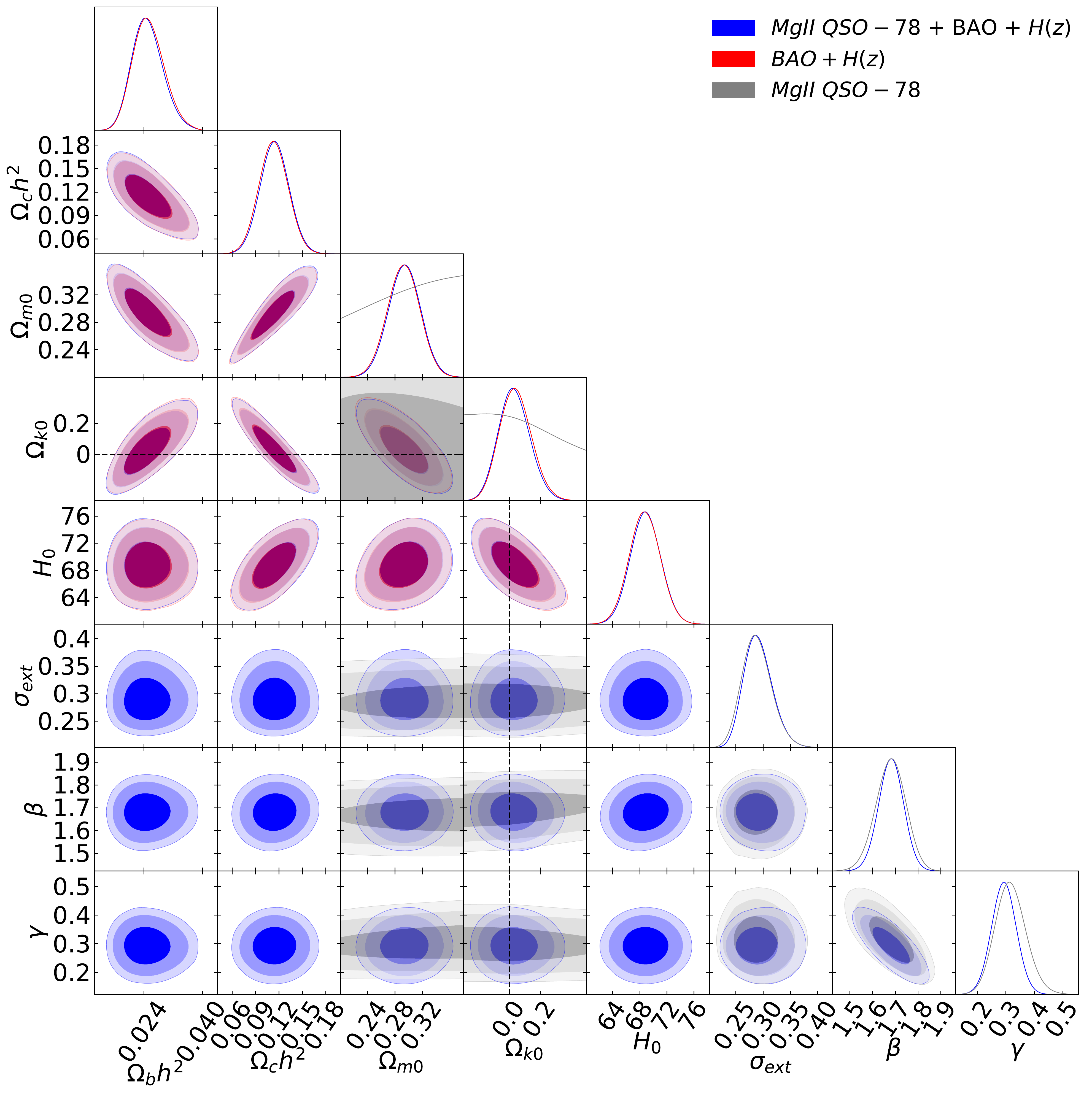}\par
\end{multicols}
\caption{One-dimensional likelihood distributions and two-dimensional likelihood contours at 1$\sigma$, 2$\sigma$, and 3$\sigma$ confidence levels using Mg II QSO-78 (gray), BAO + $H(z)$ (red), and Mg II QSO-78 + BAO + $H(z)$ (blue) data for all free parameters. Left panel shows the flat $\Lambda$CDM model and right panel shows the non-flat $\Lambda$CDM model. The black dashed straight lines in the right panel correspond to $\Omega_{k0} = 0$.}
\label{fig:Eiso-Ep}
\end{figure*}

\begin{figure*}
\begin{multicols}{2}
    \includegraphics[width=\linewidth]{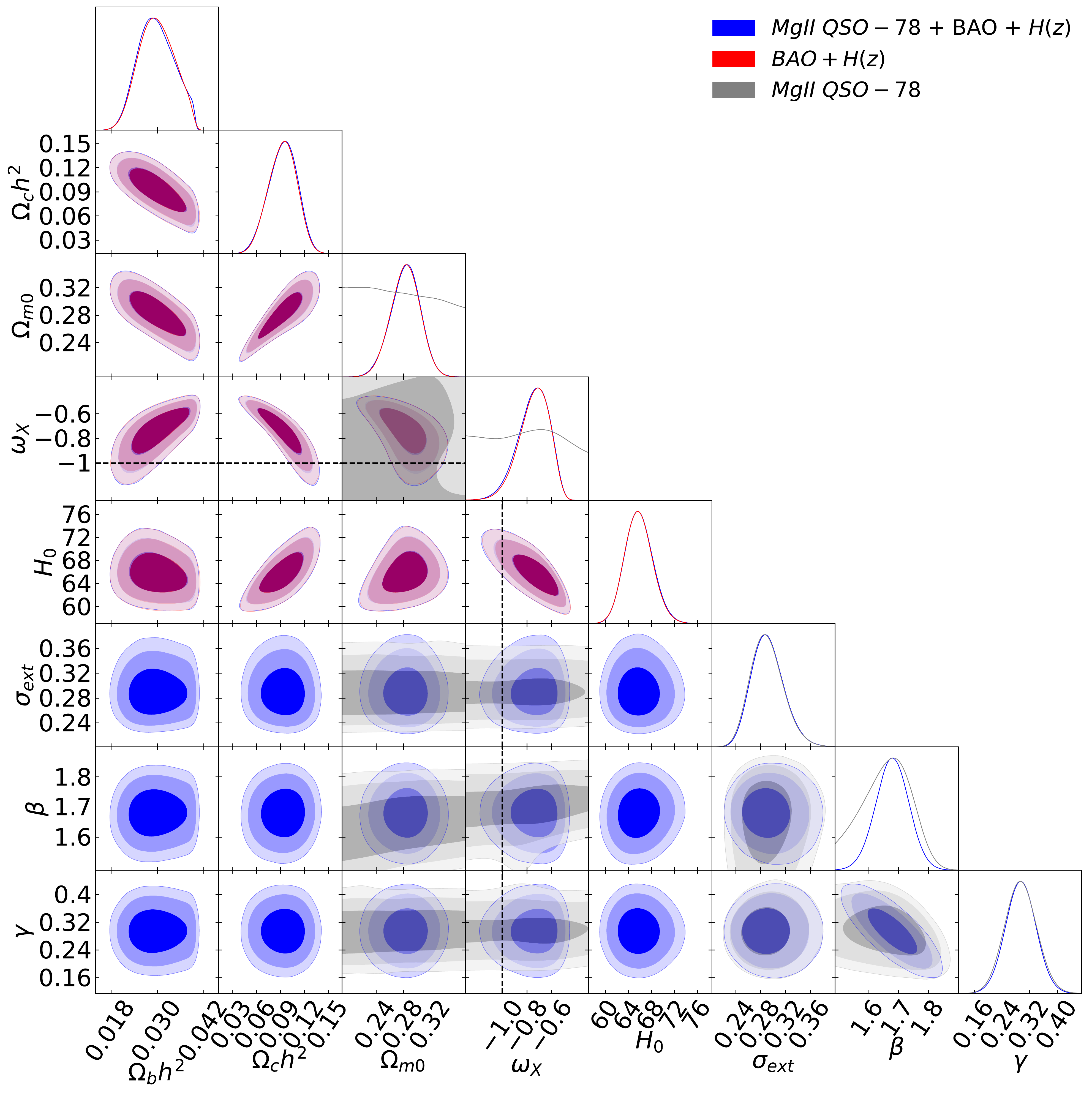}\par
    \includegraphics[width=\linewidth]{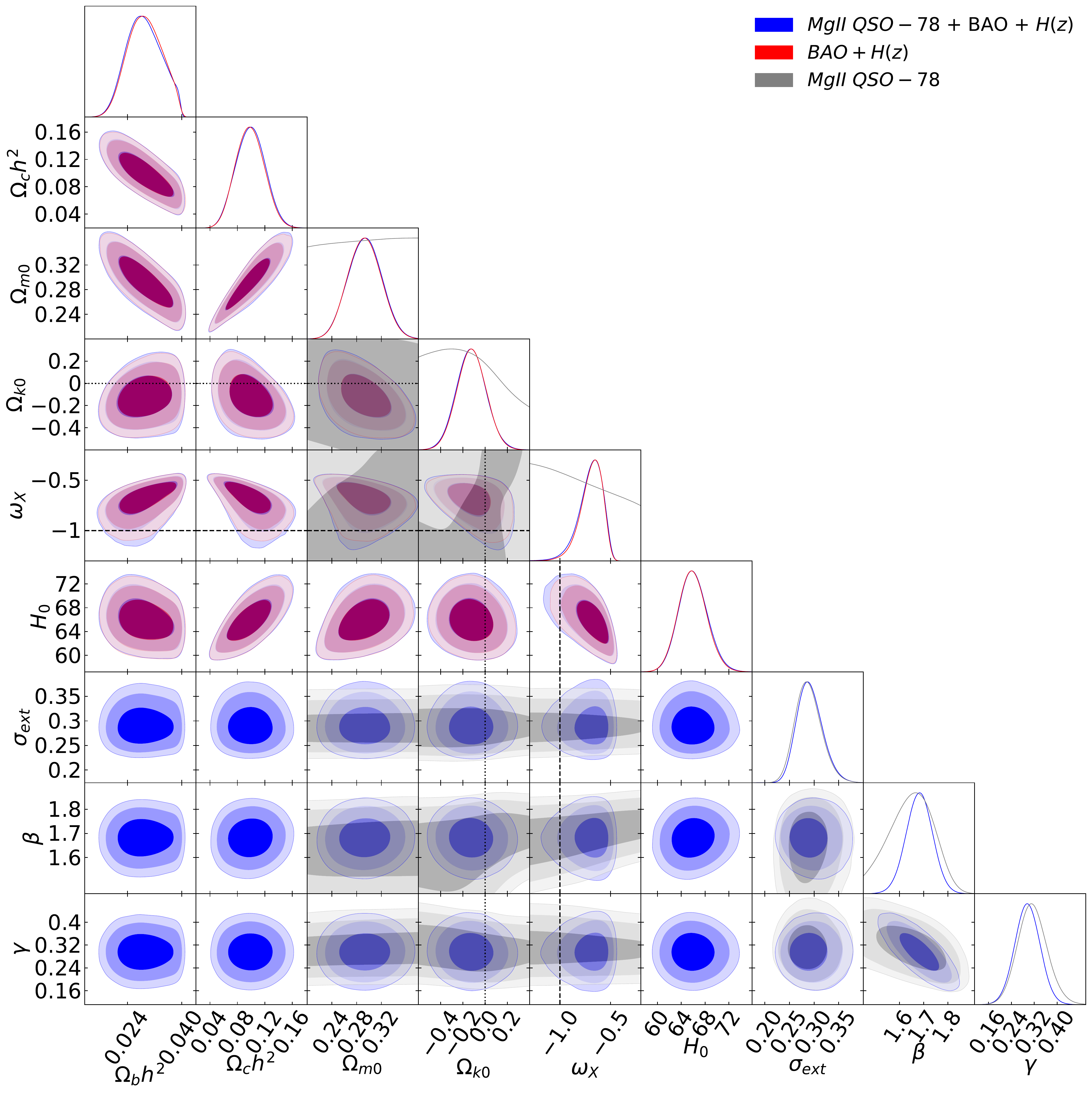}\par
\end{multicols}
\caption{One-dimensional likelihood distributions and two-dimensional likelihood contours at 1$\sigma$, 2$\sigma$, and 3$\sigma$ confidence levels using Mg II QSO-78 (gray), BAO + $H(z)$ (red), and Mg II QSO-78 + BAO + $H(z)$ (blue) data for all free parameters. Left panel shows the flat XCDM parametrization. Right panel shows the non-flat XCDM parametrization. The black dashed straight lines in both panels correspond to the $\omega_X = -1$ $\Lambda$CDM models. The black dotted straight lines in the $\Omega_{k0}$ subpanels in the right panel correspond to $\Omega_{k0} = 0$.}
\label{fig:Eiso-Ep}
\end{figure*}

\begin{figure*}
\begin{multicols}{2}
    \includegraphics[width=\linewidth]{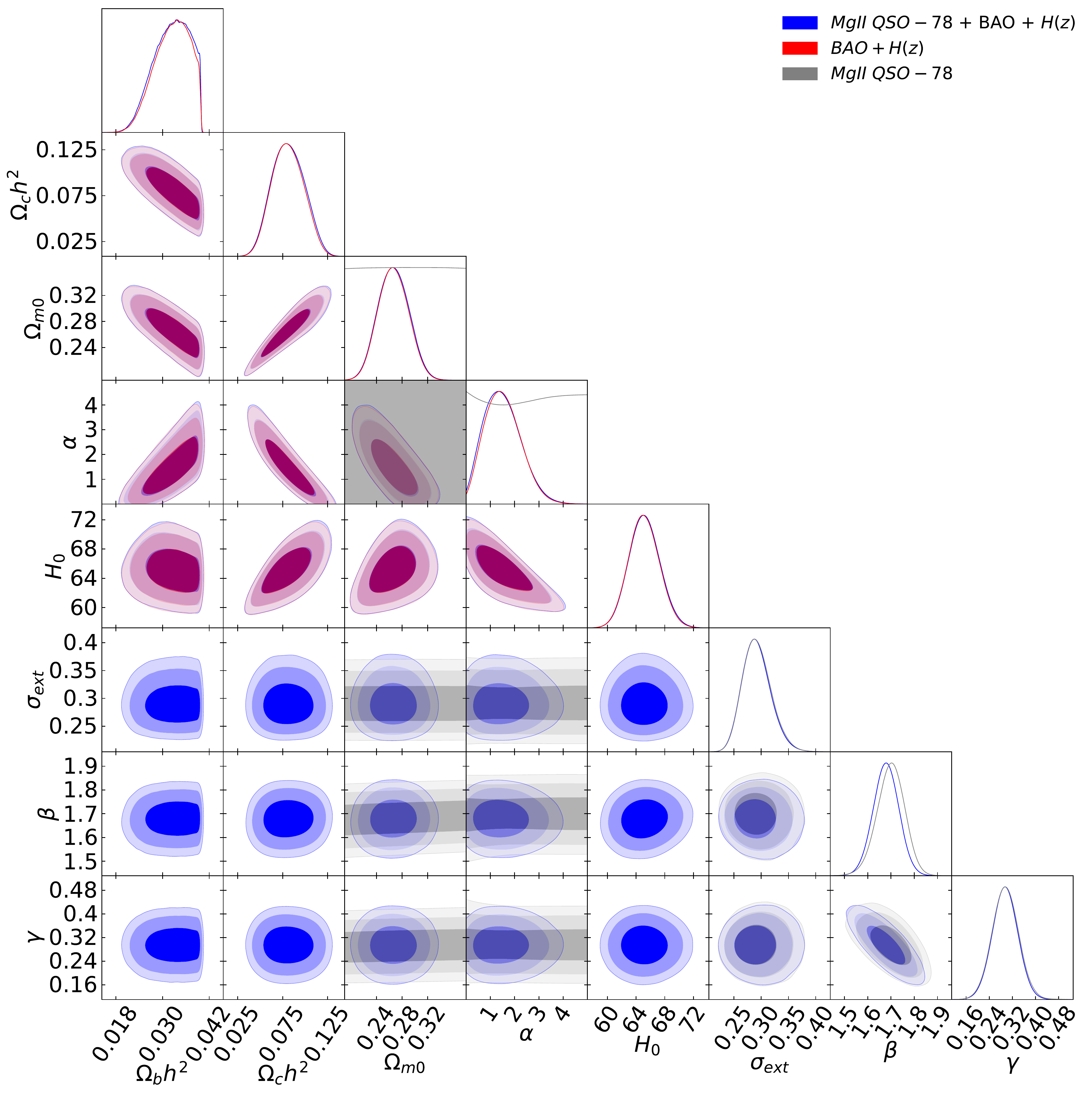}\par
    \includegraphics[width=\linewidth]{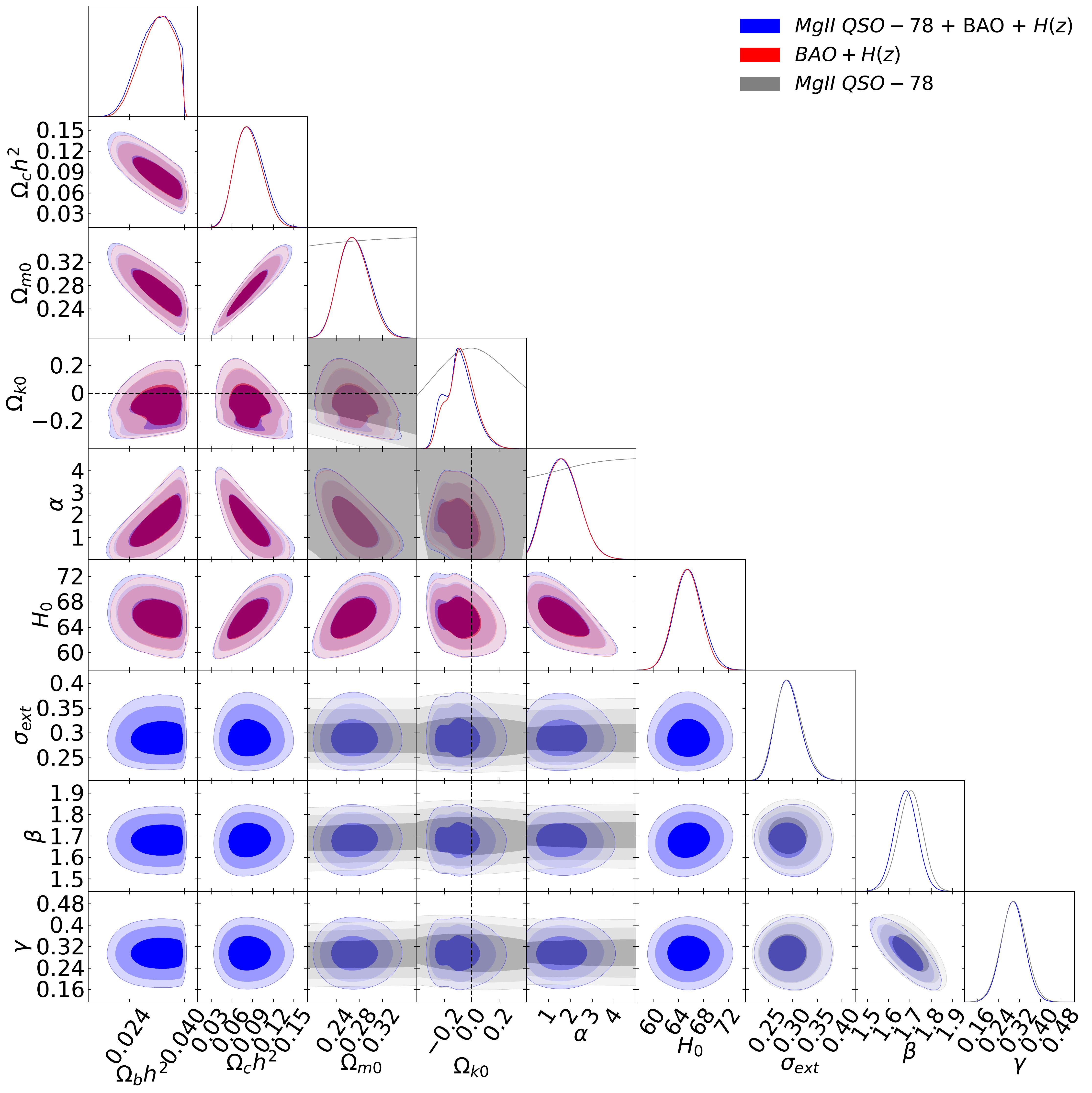}\par
\end{multicols}
\caption{One-dimensional likelihood distributions and two-dimensional likelihood contours at 1$\sigma$, 2$\sigma$, and 3$\sigma$ confidence levels using Mg II QSO-78 (gray), BAO + $H(z)$ (red), and Mg II QSO-78 + BAO + $H(z)$ (blue) data for all free parameters. Left panel shows the flat $\phi$CDM model and right panel shows the non-flat $\phi$CDM model. The $\alpha = 0$ axes correspond to the $\Lambda$CDM models. The black dashed straight lines in the $\Omega_{k0}$ subpanels in the right panel correspond to $\Omega_{k0} = 0$.}
\label{fig:Eiso-Ep}
\end{figure*}

The BAO + $H(z)$ data results listed in Tables \ref{tab:BFP} and \ref{tab:1d_BFP2} are from \cite{KhadkaRatra2021a} and are discussed in Sec.\ 5.3 of that paper. These BAO + $H(z)$ results are shown in red in Figs.\ 2--7 and 9--11. In this paper, we use these BAO + $H(z)$ results to compare with cosmological constraints obtained from the Mg II QSO data sets to see whether the Mg II QSO results are consistent or not with the BAO + $H(z)$ ones. This provides us with a qualitative idea of the consistency (inconsistency) between the Mg II QSO results and those obtained using better-established cosmological probes which favor $\Omega_{m0}\sim 0.3$.

In Figs.\ 2--4 we see that the cosmological constraints from QSO-78 data and those from BAO + $H(z)$ data are mutually consistent. It is therefore not unreasonable to jointly analyze these data. Since the Mg II QSO-78 data cosmological constraints are significantly less restrictive than those that follow from BAO + $H(z)$ data, adding the QSO-78 data to the mix will not significantly tighten the BAO + $H(z)$ cosmological constraints. Results from the Mg II QSO-78 + BAO + $H(z)$ data set are given in Tables \ref{tab:BFP} and \ref{tab:1d_BFP2}. The unmarginalized best-fit parameter values are listed in Table \ref{tab:BFP} and the one-dimensional marginalized best-fit parameter values and limits are given in Table \ref{tab:1d_BFP2}. Corresponding one-dimensional likelihood distributions and two-dimensional likelihood contours are plotted in blue in Figs.\ 9--11.

From Table \ref{tab:1d_BFP2}, the minimum value of $\Omega_bh^2$ is found to be $0.024^{+0.003}_{-0.003}$ in the spatially-flat $\Lambda$CDM model while the maximum value of $\Omega_bh^2$ is $0.032^{+0.007}_{-0.004}$ in the spatially non-flat $\phi$CDM model. The minimum value of $\Omega_ch^2$ is $0.081^{+0.018}_{-0.018}$ and is obtained in the spatially-flat $\phi$CDM model while the maximum value of $\Omega_bh^2$ is found to be $0.119^{+0.007}_{-0.008}$ in the spatially-flat $\Lambda$CDM model. The minimum value of $\Omega_{m0}$ is $0.266^{+0.024}_{-0.024}$ in the spatially-flat $\phi$CDM model and the maximum value of $\Omega_{m0}$ is $0.299^{+0.015}_{-0.017}$ in the spatially-flat $\Lambda$CDM model. As expected, these results are almost identical to those obtained using BAO + $H(z)$ data.

From Table \ref{tab:1d_BFP2}, in the flat $\Lambda$CDM model, the value of $\Omega_{\Lambda}$ is $0.700^{+0.017}_{-0.015}$. In the non-flat $\Lambda$CDM model, the value of $\Omega_{\Lambda}$ is $0.675^{+0.092}_{-0.079}$.

For analyses that involve the BAO + $H(z)$ data, the Hubble constant $H_0$ is a free parameter. From the Mg II QSO-78 + BAO + $H(z)$ data, the minimum value of $H_0$ is $65.2 \pm 2.1$ ${\rm km}\hspace{1mm}{\rm s}^{-1}{\rm Mpc}^{-1}$ in the spatially-flat $\phi$CDM model while the maximum value of $H_0$ is $69.3 \pm 1.8$ ${\rm km}\hspace{1mm}{\rm s}^{-1}{\rm Mpc}^{-1}$ in the spatially-flat $\Lambda$CDM model.

From Table \ref{tab:1d_BFP2}, the values of the spatial curvature energy density parameter $\Omega_{k0}$ are $0.031^{+0.094}_{-0.110}$, $-0.120^{+0.130}_{-0.130}$, and $-0.090^{+0.100}_{-0.120}$ in the non-flat $\Lambda$CDM, XCDM, and $\phi$CDM model respectively. These are consistent with flat spatial hypersurfaces and also with mildly open or closed ones.

From Table \ref{tab:1d_BFP2}, in the flat XCDM parametrization, the value of the dynamical dark energy equation of state parameter ($\omega_X$) is $-0.750^{+0.150}_{-0.100}$ while in the non-flat XCDM parametrization $\omega_X$ is $-0.700^{+0.140}_{-0.079}$. In the flat $\phi$CDM model, the scalar field potential energy density parameter $(\alpha)$ is $1.510^{+0.620}_{-0.890}$ while in the non-flat $\phi$CDM model $\alpha$ is $1.660^{+0.670}_{-0.850}$. In these four dynamical dark energy models, dynamical dark energy is favored at $1.7\sigma - 3.8\sigma$ statistical significance over the cosmological constant.

From Table \ref{tab:BFP}, from the $AIC$ and $BIC$ values, the most favored model is flat $\phi$CDM while non-flat $\Lambda$CDM is least favored. From the $\Delta AIC$ values, all models are almost indistinguishable from the spatially-flat $\Lambda$CDM model. From the $\Delta BIC$ values, the non-flat $\Lambda$CDM, XCDM, and $\phi$CDM models provide positive evidence for the spatially-flat $\Lambda$CDM model.

\section{Conclusion}
\label{con}

In this paper, we use the $R-L$ relation to standardize Mg II QSOs. Analyses of different Mg II QSO data sets using six different cosmological dark energy models show that the $R-L$ relation parameters are model-independent and that the intrinsic dispersion of the $R-L$ relation for the whole Mg II QSO data set is $\sim 0.29$ dex which is not very large for only 78 QSOs. So, for the first time, we have shown that one can use the $R-L$ relation to standardize available Mg II QSOs and thus use them as a cosmological probe.

We determined constraints on cosmological model parameters using these Mg II QSO data and found that these constraints are significantly weaker than, and consistent with, those obtained using BAO + $H(z)$ data. In Fig.\ \ref{fig:Hubble_diagram} we show that the 78 Mg II QSOs have a Hubble diagram consistent with what is expected in the standard spatially-flat $\Lambda$CDM model with $\Omega_{m0} = 0.3$. This differs from the results of the QSO X-ray and UV flux data compiled by \citet{RisalitiLusso2019} and \citet{Lussoetal2020}.\footnote{\citet{KhadkaRatra2021a, KhadkaRatra2021b} found that only about half of the \citet{Lussoetal2020} QSO flux sources, about a 1000 QSOs at $z \lesssim 1.5$, were standardizable and that cosmological constraints from these QSOs were consistent with what is expected in the standard $\Lambda$CDM model.} 

The constraints obtained from the joint analyses of Mg II QSO data and the BAO + $H(z)$ measurements are consistent with the current standard spatially-flat $\Lambda$CDM model but also do not rule out slight spatial curvature. These data weakly favor dynamical dark energy over the cosmological constant.

The current Mg II QSO data set contains only 78 sources and covers the redshift range $0.0033 \leq z \leq 1.89$. Future detections of significant time-delays of the BLR emission of Mg II QSOs will increase the number of sources over a larger redshift extent, which will further constrain the Mg II QSO $R-L$ relation, in particular its slope. A large increase of suitable sources is expected from the Rubin Observatory Legacy Survey of Space and Time that will monitor about 10 million quasars in six photometric bands during its 10-year lifetime. We hope that such an improved data set will soon provide tighter cosmological constraints, as well as allow for a comparison with constraints from QSO X-ray and UV flux measurements which currently are exhibiting some tension with standard flat $\Lambda$CDM model expectations.

\section{ACKNOWLEDGEMENTS}
This research was supported in part by US DOE grant DE-SC0011840, US NSF grant No.\ 161553, by the Polish Funding Agency National Science Centre, project 2017/26/A/ST9/00756 (MAESTRO 9), and by GA\v{C}R EXPRO grant 21-13491X. Part of the computation for this project was performed on the Beocat Research Cluster at Kansas State University. Time delays for quasars CTS C30.10, HE 0413-4031, and HE 0435-4312 were determined with the SALT telescope, and  Polish  participation in SALT is funded by grant No.\ MNiSW DIR/WK/2016/07. 

\section*{Data availability}
The data analysed in this article are listed in Table \ref{tab:MgQSOdata} of this paper.



\bibliographystyle{mnras}
\bibliography{mybibfile}




\onecolumn
\begin{appendix}
\section{Mg II QSO data}
\label{sec:appendix}
\addtolength{\tabcolsep}{0pt}
\LTcapwidth=\linewidth
\begin{longtable}{lccc}
\caption{Reverberation-mapped Mg II QSO samples. For each source, columns list: QSO name, redshift, continuum flux density at 3000\,\AA, and measured rest-frame time delay. The first 68 sources are from \citet{Mary2020}, the source in boldface is from \citet{Michal2021}, and the last 9 sources are from \citet{Zhefu2021}.}
\label{tab:MgQSOdata}\\
\hline\hline
Object &  $z$ &  $\log \left(F_{3000}/{\rm erg}\,{\rm s^{-1}}{\rm cm^{-2}}\right)$  &  $\tau$ (day)\\
\hline
\endfirsthead
\caption{continued.}\\
\hline\hline
Object &  $z$ &  $\log \left(F_{3000}/{\rm erg}\,{\rm s^{-1}}\,{\rm cm^{-2}}\right)$  &  $\tau$ (day)\\
\hline
\endhead
\hline
\endfoot
018& 0.848 & $-13.1412\pm0.0009$& $125.9^{+6.8}_{-7.0}$\\
028& 1.392& $-12.4734\pm0.0004$& $65.7^{+24.8}_{-14.2}$\\
038& 1.383& $-12.3664\pm0.0003$& $120.7^{+27.9}_{-28.7}$\\
044& 1.233& $-13.04308\pm0.0013$& $65.8^{+18.8}_{-4.8}$\\
102& 0.861& $-12.5575\pm0.0005$& $86.9^{+16.2}_{-13.3}$\\
114& 1.226& $-11.8369\pm0.0003$& $186.6^{+20.3}_{-15.4}$\\
118& 0.715& $-12.2592\pm0.0006$& $102.2^{+27.0}_{-19.5}$\\
123& 0.891& $-12.8942\pm0.0009$& $81.6^{+28.0}_{-26.6}$\\
135& 1.315& $-12.8122\pm0.0005$& $93.0^{+9.6}_{-9.8}$\\
158& 1.478& $-13.2376\pm0.0012$& $119.1^{+4.0}_{-11.8}$\\
159& 1.587& $-12.7139\pm0.0006$& $324.2^{+25.3}_{-19.4}$\\
160& 0.36 & $-12.8441\pm0.0013$& $106.5^{+18.2}_{-16.6}$\\
170& 1.163& $-12.6802\pm0.0005$& $98.5^{+6.7}_{-17.7}$\\
185& 0.987& $-12.8039\pm0.0094$& $387.9^{+3.3}_{-3.0}$\\
191& 0.442& $-13.0544\pm0.0012$& $93.9^{+24.3}_{-29.1}$\\
228& 1.264& $-13.2697\pm0.0011$& $37.9^{+14.4}_{-9.1}$\\
232& 0.808& $-13.1895\pm0.0014$& $273.8^{+5.1}_{-4.1}$\\
240& 0.762& $-13.3270\pm0.0021$& $17.2^{+3.5}_{-2.8}$\\
260& 0.995& $-12.4126\pm0.0004$& $94.9^{+18.7}_{-17.2}$\\
280& 1.366& $-12.5531\pm0.0003$& $99.1^{+3.3}_{-9.5}$\\
285& 1.034& $-13.2539\pm0.0020$& $138.5^{+15.2}_{-21.1}$\\
291& 0.532& $-13.2471\pm0.0016$& $39.7^{+4.2}_{-2.6}$\\
294& 1.215& $-12.4272\pm0.0004$& $71.8^{+17.8}_{-9.5}$\\
301& 0.548& $-12.8782\pm0.0011$& $136.3^{+17.0}_{-16.9}$\\
303& 0.821& $-13.3066\pm0.0013$& $57.7^{+10.5}_{-8.3}$\\
329& 0.721& $-11.9680\pm0.0007$& $87.5^{+23.8}_{-14.0}$\\
338& 0.418& $-12.9969\pm0.0013$& $22.1^{+8.8}_{-6.2}$\\
419& 1.272& $-12.9765\pm0.0011$& $95.5^{+15.2}_{-15.5}$\\
422& 1.074& $-13.0946\pm0.0011$& $109.3^{+25.4}_{-29.6}$\\
440& 0.754& $-12.5157\pm0.0004$& $114.6^{+7.4}_{-10.8}$\\
441& 1.397& $-12.5772\pm0.0004$& $127.7^{+5.7}_{-7.3}$\\
449& 1.218& $-12.9299\pm0.0013$& $119.8^{+14.7}_{-24.4}$\\
457& 0.604& $-13.4805\pm0.0029$& $20.50^{+7.7}_{-5.3}$\\
459& 1.156& $-12.8737\pm0.0011$& $122.8^{+5.1}_{-5.7}$\\
469& 1.004& $-12.1222\pm0.0002$& $224.1^{+27.9}_{-74.3}$\\
492& 0.964& $-12.3786\pm0.0004$& $92.0^{+16.3}_{-12.7}$\\
493& 1.592& $-12.2173\pm0.0004$& $315.6^{+30.7}_{-35.7}$\\
501& 1.155& $-12.9728\pm0.0009$& $44.9^{+11.7}_{-10.4}$\\
505& 1.144& $-13.0625\pm0.0011$& $94.7^{+10.8}_{-16.7}$\\
522& 1.384& $-12.9671\pm0.0006$& $115.8^{+11.3}_{-16.0}$\\
556& 1.494& $-12.6492\pm0.0005$& $98.7^{+13.9}_{-10.8}$\\
588& 0.998& $-12.1158\pm0.0002$& $74.3^{+23.0}_{-18.2}$\\
593& 0.992& $-12.7093\pm0.0006$& $80.1^{+21.4}_{-20.8}$\\
622& 0.572& $-12.6232\pm0.0005$& $61.7^{+6.0}_{-4.3}$\\
645& 0.474& $-12.7268\pm0.0009$& $30.2^{+26.8}_{-8.9}$\\
649& 0.85 & $-13.0437\pm0.0013$& $165.5^{+22.2}_{-25.1}$\\
651& 1.486& $-12.9434\pm0.0011$& $76.5^{+18.0}_{-15.6}$\\
675& 0.919& $-12.5273\pm0.0005$& $139.8^{+12.0}_{-22.6}$\\
678& 1.463& $-12.8267\pm0.0007$& $82.9^{+11.9}_{-10.2}$\\
709& 1.251& $-12.9586\pm0.0010$& $85.4^{+17.7}_{-19.3}$\\
714& 0.921& $-12.8296\pm0.0012$& $320.1^{+11.3}_{-11.2}$\\
756& 0.852& $-13.1462\pm0.0023$& $315.3^{+20.5}_{-16.4}$\\
761& 0.771& $-12.6395\pm0.0024$& $102.1^{+8.2}_{-7.4}$\\
771& 1.492& $-12.4477\pm0.0004$& $31.3^{+8.1}_{-4.6}$\\
774& 1.686& $-12.5786\pm0.0004$& $58.9^{+13.7}_{-10.1}$\\
792& 0.526& $-13.5353\pm0.0030$& $111.4^{+29.5}_{-20.0}$\\
848& 0.757& $-13.3199\pm0.0015$& $65.1^{+29.4}_{-16.3}$\\
J141214& 0.4581& $-12.2526\pm0.00043$& $36.7^{+10.4}_{-4.8}$\\
J141018& 0.4696& $-13.1883\pm0.00506$& $32.3^{+12.9}_{-5.3}$\\
J141417& 0.6037& $-13.4926\pm0.0029$& $29.1^{+3.6}_{-8.8}$\\
J142049& 0.751 & $-12.7205\pm0.0009$& $34.0^{+6.7}_{-12.0}$\\
J141650& 0.5266& $-13.2586\pm0.00198$& $25.1^{+2.0}_{-2.6}$\\
J141644& 0.4253& $-12.8667\pm0.00105$& $17.2^{+2.7}_{-2.7}$\\
CTS252& 1.89 &  $-11.6068\pm0.09142$& $190.0^{+59.0}_{-114.0}$\\
NGC4151& 0.0033& $-9.5484\pm0.18206$& $6.8^{+1.7}_{-2.1}$\\
NGC4151& 0.0033& $-9.5484\pm0.18206$& $5.3^{+1.9}_{-1.8}$\\
CTSC30& 0.9005& $-11.5825\pm0.026$& $564.0^{+109.0}_{-71.0}$\\
HE0413-4031& 1.3765& $-11.3203\pm0.0434$& $302.9^{+23.7}_{-19.1}$\\
\textbf{HE0435-4312}& 1.2231& $-11.5754\pm0.036$& $296^{+13.0}_{-14.0}$\\
J025225.52+003405.90& 1.62425& $-12.6489\pm0.05203$& $198.82^{+16.96}_{-19.03}$\\
J021612.83-044634.10& 1.56043& $-12.7064\pm0.04075$& $51.46^{+14.37}_{-8.95}$\\
J033553.51-275044.70& 1.57774& $-12.3688\pm0.04223$& $48.14^{+22.05}_{-8.82}$\\
J003710.86-444048.11& 1.06703& $-12.1225\pm0.04395$& $191.76^{+27.62}_{-18.47}$\\
J003207.44-433049.00& 1.53278& $-12.5873\pm0.02829$& $146.97^{+2.43}_{-0.87}$\\
J003015.00-430333.52& 1.64984& $-12.7328\pm0.04608$& $185.55^{+14.55}_{-4.72}$\\
J003052.76-430301.08& 1.42754& $-12.5959\pm0.03380$& $166.76^{+11.00}_{-10.88}$\\
J003234.33-431937.81& 1.64058& $-12.5643\pm0.03011$& $248.82^{+18.10}_{-11.64}$\\
J003206.50-425325.22& 1.7496 & $-12.7498\pm0.09277$& $157.80^{+12.77}_{-4.95}$\\
\hline
\end{longtable}
\end{appendix}



\bsp	
\label{lastpage}
\end{document}